\documentclass[twocolumn]{aastex63}

\usepackage{aas_macros}

\usepackage{color}
\usepackage{listings}
\usepackage{subfigure}

\usepackage{amsmath}
\usepackage{xspace}
\usepackage{xcolor}

\newcommand{\kms}         {km\,s$^{-1}$} 

\newcommand{\msun}        {$M_{\odot}$}

\def\spose#1{\hbox to 0pt{#1\hss}}
\def\lta{\mathrel{\spose{\lower 3pt\hbox{$\mathchar"218$}}
     \raise 2.0pt\hbox{$\mathchar"13C$}}}
\def\gta{\mathrel{\spose{\lower 3pt\hbox{$\mathchar"218$}}
    \raise 2.0pt\hbox{$\mathchar"13E$}}}

\shorttitle{Orbital Solutions for Gaia DR3 BH and NS Candidates}
\shortauthors{Simon et al.}

\begin{document}

\title{Radial Velocity Orbital Solutions for Candidate Black Hole and
  Neutron Star Binary Systems in the Gaia Data Release 3
  Catalog\footnote{This paper includes data gathered with the 6.5
    meter Magellan Telescopes located at Las Campanas Observatory,
    Chile.}}

\author{Joshua~D.~Simon}
\affiliation{Observatories of the Carnegie Institution for Science, 813 Santa Barbara St., Pasadena, CA 91101, USA}

\author{Casey~Y.~Lam}
\affiliation{Observatories of the Carnegie Institution for Science, 813 Santa Barbara St., Pasadena, CA 91101, USA}

\author{Kareem~El-Badry}
\affiliation{Department of Astronomy, California Institute of Technology, 1200 E. California Blvd., Pasadena, CA 91125, USA}

\author{Henrique Reggiani}
\affiliation{Gemini South, Gemini Observatory, NSF’s NOIRLab, Casilla 603, La Serena, Chile}

\author{Sukanya Chakrabarti}
\affiliation{Department of Physics and Astronomy, University of Alabama, Huntsville, Huntsville, Alabama 35899, USA}

\author{Puragra Guhathakurta}
\affiliation{Department of Astronomy and Astrophysics, University of California Santa Cruz, 1156 High Street, Santa Cruz, CA 95064, USA}

\author{Ian B. Thompson}
\affiliation{Observatories of the Carnegie Institution for Science, 813 Santa Barbara St., Pasadena, CA 91101, USA}

\author{Nidia Morrell}
\affiliation{Las Campanas Observatory, Carnegie Observatories, Casilla 601, La Serena, Chile}

\author{Daniel Huber}
\affiliation{Institute for Astronomy, University of Hawaii at Mānoa, 2680 Woodlawn Drive, Honolulu, HI 96822, USA}

\author{Benjamin J. Fulton}
\affiliation{NASA Exoplanet Science Institute, IPAC, MS 100-22, Caltech, 1200 E. California Blvd, Pasadena, CA 91125, USA}

\author{Lauren M. Weiss}
\affiliation{Department of Physics and Astronomy, University of Notre Dame, Notre Dame, IN 46556, USA}

\begin{abstract}
We present spectroscopic followup observations of binary systems from
the Gaia Data Release 3 (DR3) binary catalog that were selected to
have large enough mass functions for their companions to be black
holes or neutron stars.  The selection includes 20 stars that are
astrometric and/or spectroscopic binaries, as well as 11 stars with
large accelerations both in the plane of the sky and along the line of
sight but no DR3 orbital solution.  We provide classifications for
this entire sample, including radial velocity orbital solutions for 12
binaries.  Apart from the previously published binaries Gaia~BH1,
Gaia~BH2, and Gaia~NS1, we show that the Gaia orbits are incorrect for
all of the stars with candidate dark companions above 2~\msun.  We
suggest more conservative cuts on the significance and goodness of fit
parameters that may be useful for identifying reliable orbital
solutions in the tail of the binary star distribution.  Although we
find no new confirmed black hole or neutron star companions, one
accelerating system has a minimum companion mass of
$1.16\pm0.01$~\msun\ that is likely to be a neutron star or an
ultramassive white dwarf.  The acceleration catalogs may therefore
provide a largely unexplored source of additional wide binaries
containing compact objects.

\end{abstract}


\section{Introduction}
\label{sec:intro}

The Milky Way is expected to host millions of stellar-mass black holes, but only a few dozen have actually been discovered
\citep[e.g.,][]{cs16}.  The properties of these black holes can shed
light on numerous open questions related to the fate of massive stars
and the evolution of binary (or multiple) star systems that contain
black holes.  Recently, the combination of the first significant
sample of merging binary black holes
\citep[e.g.,][]{abbott16,ligoO2,ligoO3a} and the availability of large
spectroscopic and astrometric surveys of binary systems
\citep[e.g.,][]{badenes18,pw20,tian20,kounkel21,halbwachs23,gosset25}
has sparked increased interest in the demographics of black hole
binaries \citep[e.g.,][]{breivik17,olejak20,chawla22,shikauchi23}.

A number of candidate binary systems containing a black hole have been
identified via analysis of spectroscopic survey data and studies of
individually interesting objects
\citep{giesers18,giesers19,liu19,thompson19,rivinius20,jayasinghe21,saracino22,lennon22}.
However, the nature of nearly all of these objects has been seriously
questioned or refuted
\citep{ebq20,eldridge20,bodensteiner20,gw20,ebb22,elbadry22a,elbadry22b,bianchi24}.
Only the binaries in the globular cluster NGC~3201
\citep{giesers18,giesers19} and two O stars with massive companions \citep{mahy22,shenar22} remain likely black hole candidates.

The astrometric and spectroscopic binary star catalogs created by the
Gaia Collaboration as part of the third data release (DR3) offer a new
route to identifying binary systems containing a black hole or neutron
star along with a luminous companion star
\citep{gaiadr3arenou,halbwachs23,gosset25}.  Two confirmed black holes
and a number of neutron stars have been identified in the DR3 orbital
solution catalog\footnote{A third black hole was discovered in pre-release DR4 data as well \citep{panuzzo24}, but this object was not included in any of the DR3 non-single stars catalogs because of its long orbital period.}
\citep{elbadry23a,elbadry23b,chakrabarti23,elbadry24a,elbadry24b}.
However, a few dozen other candidates selected from the same catalog
\citep{andrews22,gaiadr3arenou,shahaf23} either remain
unconfirmed or have been ruled out as containing a compact object
\citep{ebr22,elbadry23a}.

In this study, we investigate a nearly complete sample of candidate
neutron star or black hole binaries from either the DR3 astrometric
catalog or the spectroscopic binary catalog.  We derive radial
velocity orbital solutions for these sources in order to determine
their nature and provide insight into the failure modes of the Gaia
binary solutions for putative massive dark companions.  

In \S~\ref{sec:obs}, we define the sample and describe the observations and analysis methods.  We present our results on the properties and nature of each object in the target sample in \S~\ref{sec:results}.  In \S~\ref{sec:discussion}, we summarize our findings for the full sample and briefly examine the implications of these results for detection of black hole and neutron star binary systems in Gaia~DR4.  We present our conclusions in \S~\ref{sec:conclusion}.

\section{Observations}
\label{sec:obs}

\subsection{Sample Selection}
\label{sec:sample}

We selected candidate compact object binaries both directly from the
Gaia DR3 binary catalogs and using independent analyses of the Gaia
catalogs in the literature.  Specifically, \citet{shahaf23} listed eight
sources that they classified as astrometric mass ratio function (AMRF)
Class III with a companion mass above 2~M$_{\odot}$, for which both a
single main sequence star or a close main sequence binary are ruled
out as possible companions.  Seven of the eight stars are from the astrometric catalog, and one (Gaia~DR3~3263804373319076480) has an AstroSpectroSB1 solution.  We note that none of these systems
contain UV-bright companions identified by \citet{ganguly23}.
\citet{andrews22} identified two additional binaries with inferred
companion masses above 2~M$_{\odot}$, as well as one source in common
with \citeauthor{shahaf23} In addition, \citet{tanikawa23} found a
candidate black hole binary that was not included in either of the
other searches.  Three of these sources are Gaia~BH1
\citep{elbadry23a,chakrabarti23}, Gaia~BH2 \citep{elbadry23b}, and
Gaia~NS1 \citep{elbadry24a}.  These systems were followed up as part
of this program, but since they have been studied in detail elsewhere, we
include them for completeness without carrying out any additional
analysis here.

\subsubsection{Gaia DR3 binary catalogs}

To add to the set of neutron star (NS) and black hole candidates identified from DR3 in the literature, we also searched for sources in the Gaia binary catalogs with massive, dark companions.
In the DR3 astrometric catalog of binary masses, three sources satisfy
the criteria $P > 9$~d, $\mathrm{m2\_lower} \ge 5$~\msun\,
$\mathrm{m2\_lower} > \mathrm{m1}$, and $\mathrm{combination\_method}
\ne \mathrm{`SB2}$' (period longer than 9~d, a lower limit on the secondary mass that is above 5~\msun\ and larger than the primary mass, and not a double-lined binary).  These selection cuts are intended to identify detached binaries where the secondary is both dark and unambiguously in the black hole mass range.  Note that masses are not directly observable by Gaia, but were estimated for some classes of binaries in the DR3 catalog using the methods described by \citet{gaiadr3arenou}.  Two of the three objects selected via these criteria were also in the \citet{shahaf23}
AMRF Class~III list (Gaia~DR3~4373465352415301632 = Gaia~BH1, and
Gaia~DR3~6281177228434199296), but one additional source
(Gaia~DR3~3640889032890567040) with a perhaps implausible secondary
mass of $>79.4$~M$_{\odot}$ was not.

In the DR3 catalog of spectroscopic binaries, there are ten systems
meeting the same criteria.  From this list, we eliminated
Gaia~DR3~5824739062379517824, which is a known Be star (HD~134401),
and therefore likely a contaminant based on previous efforts to
identify black hole companions to Be star primaries
\citep[e.g.,][]{bodensteiner20,gw20,elbadry22a,janssens23,mh26},
and Gaia~DR3~2086448353089047808 and Gaia~DR3~6102598776102841344,
which exhibit eclipses in their TESS light curves and therefore
contain two luminous stars.

In the AstroSpectroSB1 catalog of binaries with both astrometric and
spectroscopic solutions, Gaia~DR3~1864406790238257536 is the only
additional object satisfying the selection criteria.  This source was not discussed by either \citet{andrews22} or \citet{shahaf23}.

\begin{figure*}
\centering
    \subfigure{\includegraphics[width=0.47\textwidth]{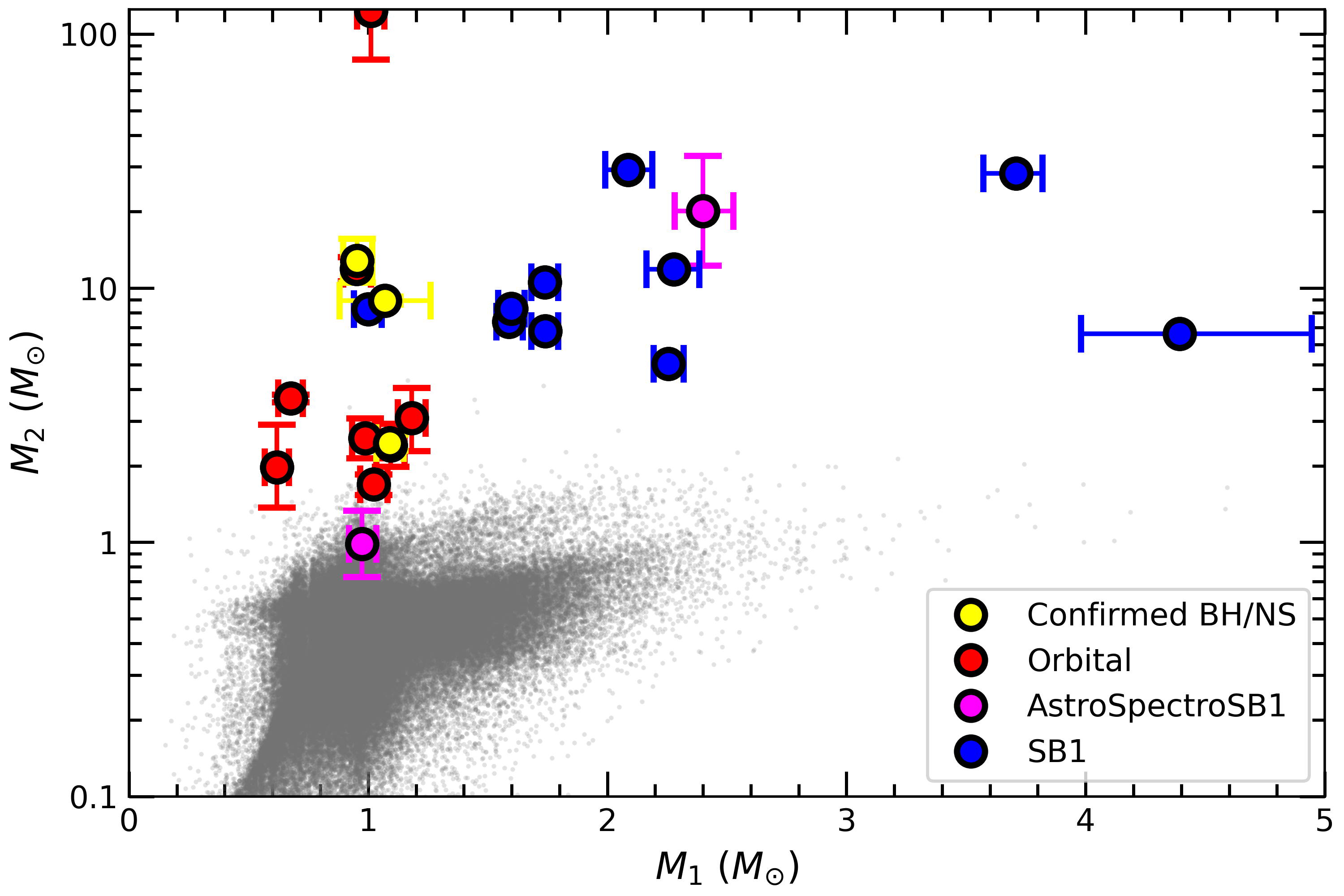}}
    \subfigure{\includegraphics[width=0.47\textwidth]{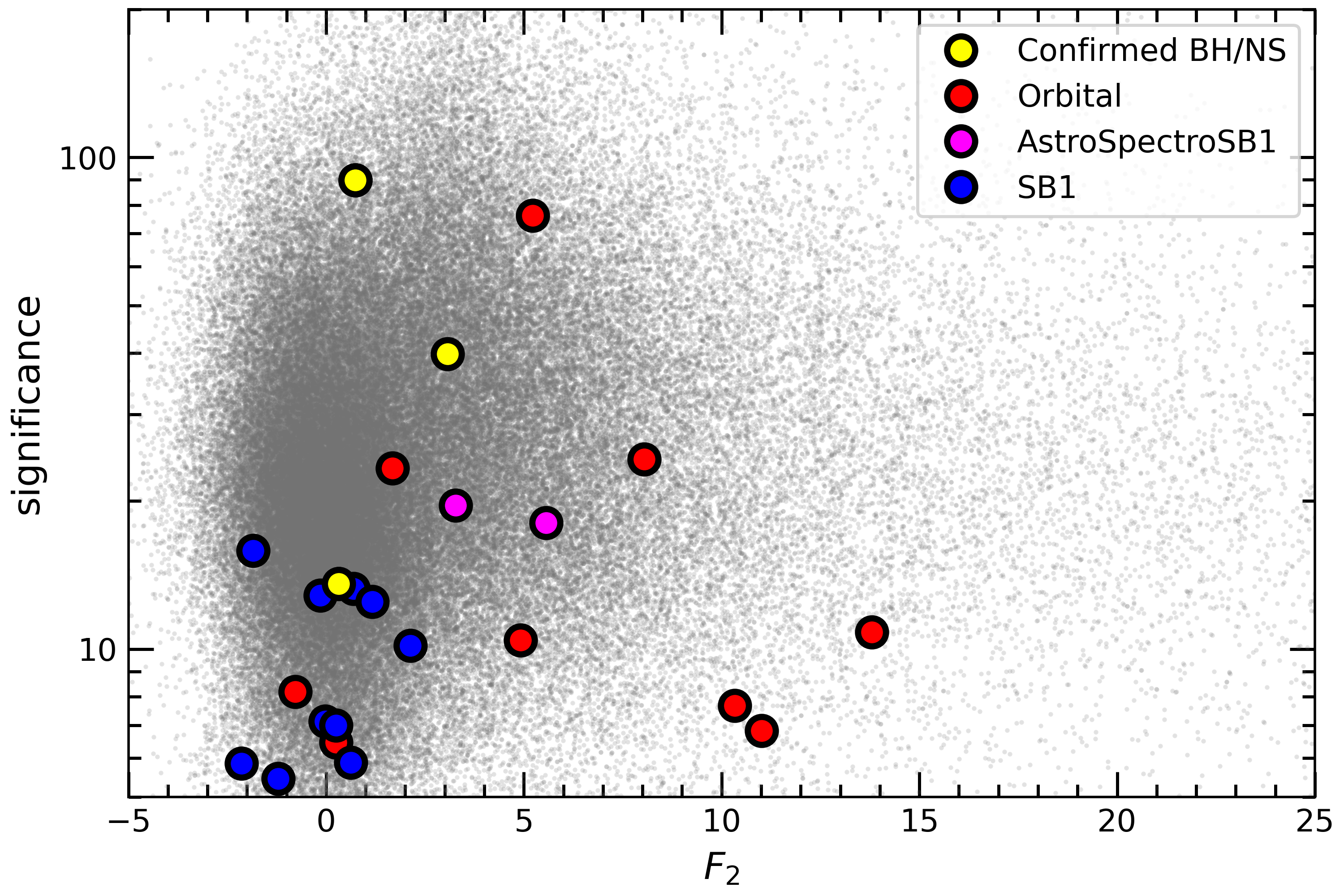}}
\vskip\baselineskip
    \subfigure{\includegraphics[width=0.47\textwidth]{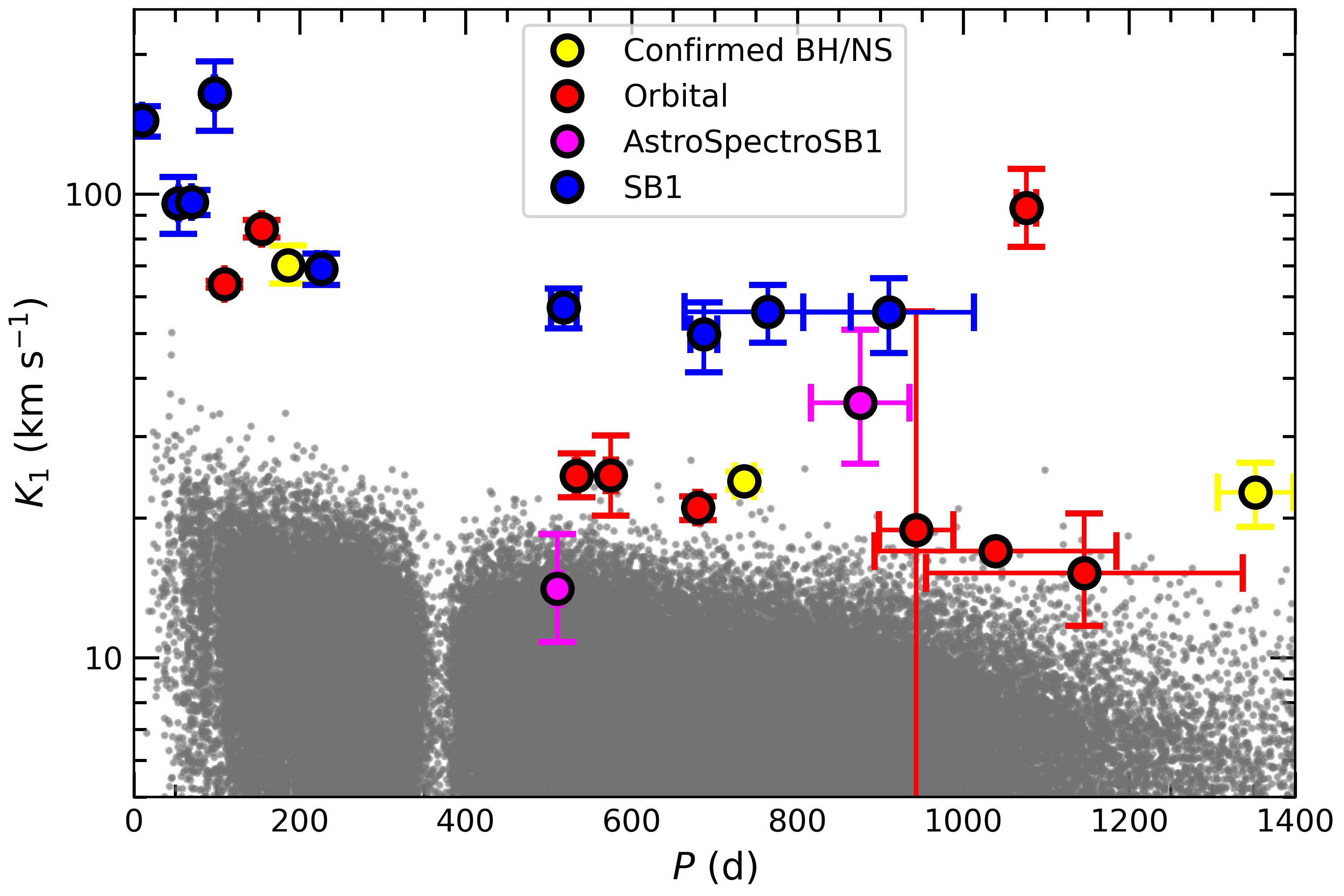}}
    \subfigure{\includegraphics[width=0.47\textwidth]{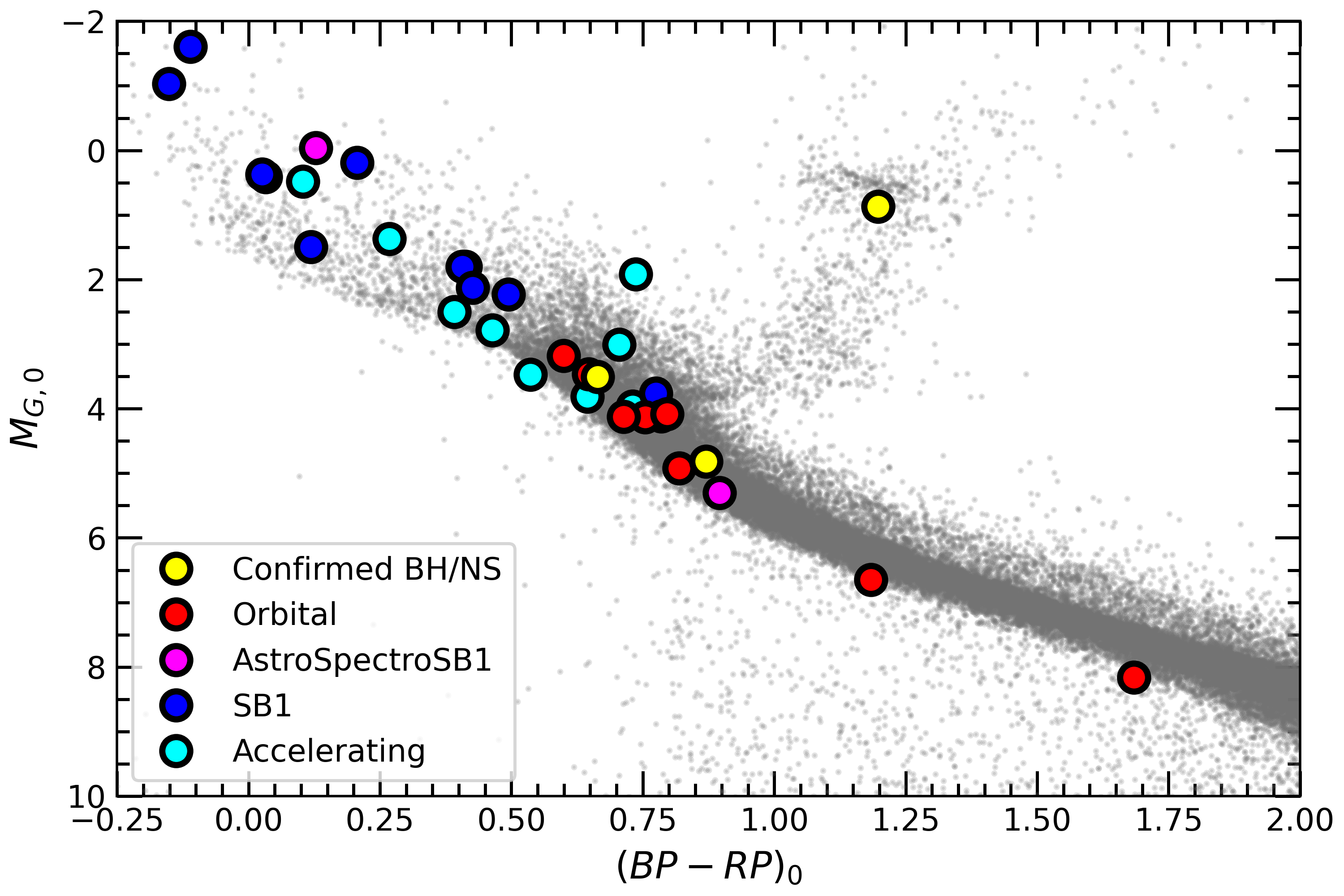}}
\caption{(upper left) Secondary mass as a function
      of primary mass for Gaia DR3 binaries.  The confirmed black hole and neutron star
      binaries Gaia~BH1, Gaia~BH2, and Gaia~NS1 are displayed as
      yellow circles.  The candidates selected from the DR3 orbital
      catalog are plotted as red circles, those from the
      AstroSpectroSB1 catalog are plotted as magenta circles, and the
      spectroscopic binary candidates are plotted as dark blue
      circles.  The stars in the DR3
      catalog of binary masses are shown as small gray dots for
      comparison.  (upper right) Orbital solution significance as a function
      of goodness of fit ($F_{2}$).  The symbols are the same as in
      the upper left panel.  Many, although not all, of the candidates
      investigated in this paper have lower significance orbits and/or
      worse goodness of fit values than the previously confirmed
      compact object binaries.  (lower left) Velocity semi-amplitudes as a function of binary period.  The colored symbols are the same as in
      the upper left panel, with stars in the DR3 catalog of binary masses shown as small gray dots for comparison.  Note
      that to compute the semi-amplitudes, we used the m2\_lower field
      in the catalog rather than m2 so that each star would have a
      value and so that spectroscopic and astrometric binaries can be
      compared on an equal footing.  (lower right) Gaia color-magnitude diagram of the candidate sample.
      The colored symbols are the same as in the upper left panel.  A randomly selected 20\% of the stars
      within 150~pc of the sun are shown as small gray dots for
      comparison.  For all stars, we dereddened the photometry using
      the 3D dust map from \citet{wang25}.
      The tendency for the SB1 binaries, and to a lesser
      extent the accelerating systems, to have hotter and more massive
      primaries is visually apparent.  Almost all of the sample is
      located close to the main sequence, such that massive
      secondaries should be easily detectable if they are luminous.}
\label{fig:sample_4panel}
\end{figure*}

Some properties of the selected binary systems and their Gaia DR3 orbital parameters are illustrated in Fig.~\ref{fig:sample_4panel}.  Our sample
covers the extreme tail of the distribution of secondary masses in the
DR3 binary catalog.  Most of the sample does not stand out in primary
mass.
Some of the stars have relatively poor fits ($F_{2} > 10$) and/or
relatively low signal-to-noise ratios ($\mathrm{significance} < 10$), but
others are located within the bulk of the DR3 distribution and with
values similar to those of, e.g., Gaia~BH1.

The basic orbital properties of the sample are shown in the lower left panel of
Fig.~\ref{fig:sample_4panel}.  The velocity semi-amplitudes ($K_{1}$) for astrometric binaries are computed from the Thiele-Innes coefficients as described by \citet{binnendijk60}.  For spectroscopic binaries, the $K_{1}$ value shown is taken directly from the DR3 catalog.  By construction, the selected binaries
represent the largest velocity semi-amplitudes at all periods.  All
binary systems with $K_{1} > 30$~\kms\ and $P > 200$~d are included in
the sample.  A few remaining systems with $20 < K_{1} < 30$~\kms\ and
periods of up to $\sim2$~yr could be targets for future follow-up
programs.

\subsubsection{Gaia DR3 acceleration catalogs}

To explore the possibility of massive companions on long-period
orbits, we also selected a sample of stars with large accelerations
from the Gaia DR3 catalogs of non-single stars exhibiting
line-of-sight and astrometric accelerations that did not receive
orbital solutions.  We used the following query to identify 11 single-lined binary stars
with significant accelerations both along the line of sight and in the
plane of the sky as well as large radial velocity amplitudes:

\begin{lstlisting}[language=SQL]
select * from 
    gaiadr3.gaia_source as gaia,
    gaiadr3.nss_acceleration_astro as naa,
    gaiadr3.nss_non_linear_spectro as nls
where
    (nls.nss_solution_type = 
        'FirstDegreeTrendSB1' or 
     nls.nss_solution_type = 
        'SecondDegreeTrendSB1') and
    gaia.phot_g_mean_mag < 14 and
    gaia.rv_amplitude_robust > 50 and
    naa.goodness_of_fit < 10 and
    gaia.source_id = naa.source_id and
    gaia.source_id eq nls.source_id.
\end{lstlisting}

\begin{figure*}[bh!]
    \centering
    \hspace{-0.25in}\includegraphics[width=0.48\textwidth]{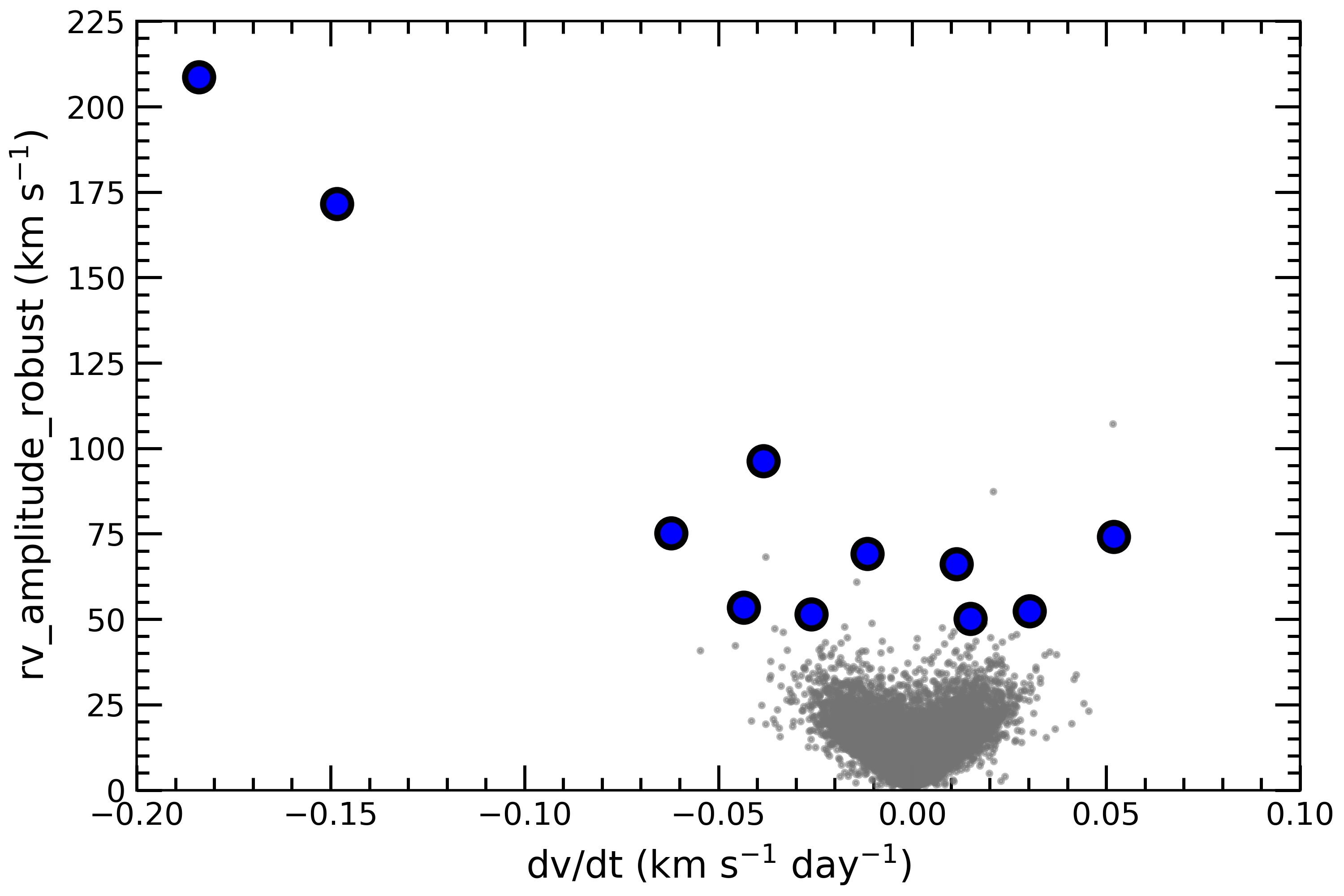}
    \hspace{0.35in}\includegraphics[width=0.48\textwidth]{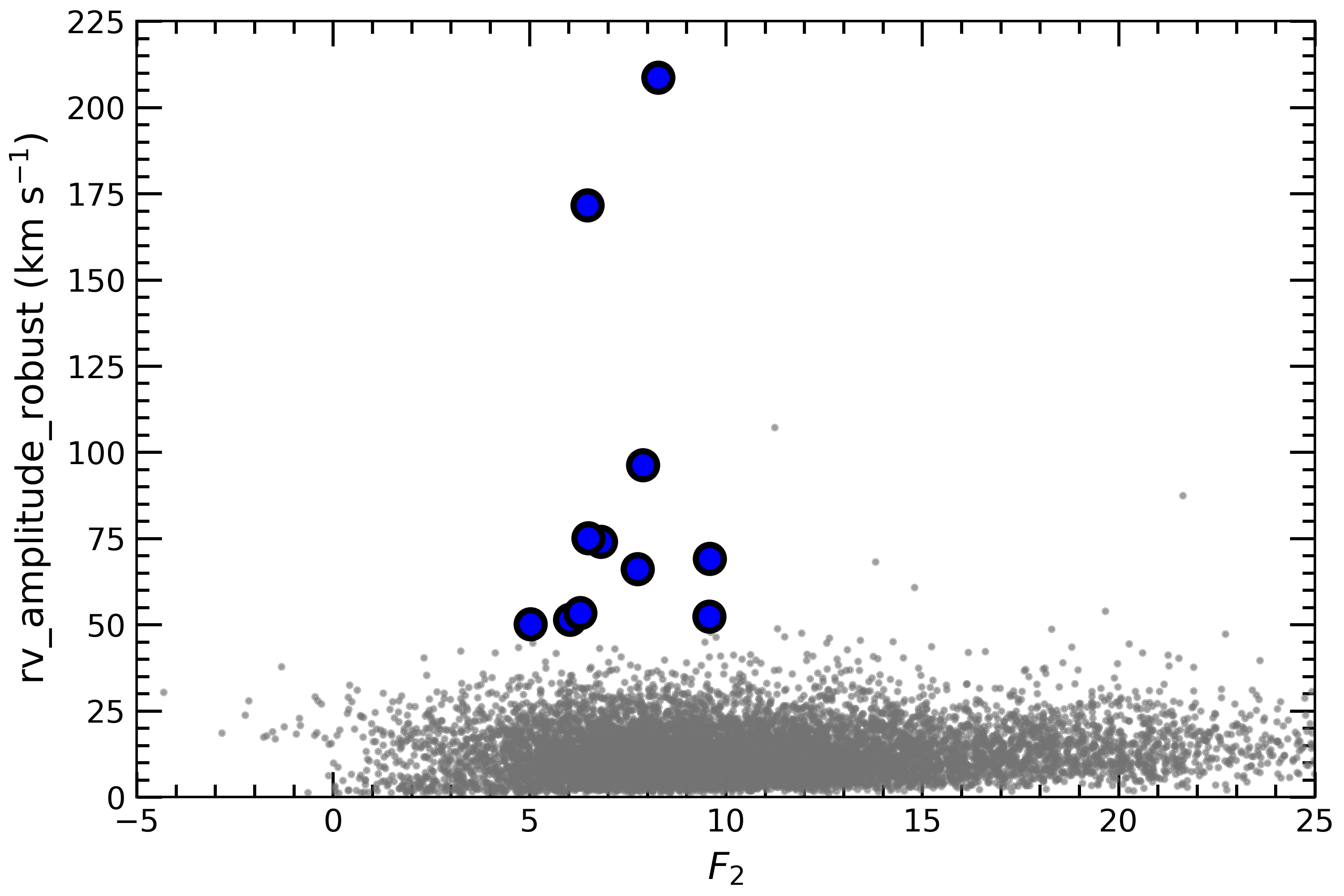}
    \caption{Properties of the acceleration sample.  (left) RV
      amplitude as a function of the first derivative of the RV.  The
      full sample of stars for which both astrometric and radial
      velocity accelerations are detected is shown in gray.  The 11 stars we
      selected for spectroscopic followup are plotted as blue circles.
      (right) RV amplitude against goodness of fit.  The symbols are
      the same as in the left panel.  There are a handful of
      additional stars with RV amplitudes above 50~\kms, but they have
      worse goodness of fit values.}
     \label{fig:accelerating_sample}
\end{figure*}

\noindent
The accelerating sources are compared to the entire group of stars
with both line-of-sight and astrometric acceleration solutions in
Fig.~\ref{fig:accelerating_sample}.

Our full sample is listed in Table~\ref{tab:sample}.  The top section of the table lists the previously-confirmed black hole and neutron star binaries, followed by astrometric binaries, spectroscopic binaries, binaries with combined astrometric and spectroscopic solutions, and accelerating sources.  The Gaia
color-magnitude diagram of the sample is displayed in the lower right panel of Fig.~\ref{fig:sample_4panel}.  Most of the selected stars are located along the main sequence, with the exception of Gaia~BH2.  A few of the AstroSpectroSB1, SB1, and accelerating sources may be slightly evolved but have not crossed the Hertzsprung gap.  The accelerating and SB1 samples have a clear tendency to contain hotter and more massive primary stars, and we note here that hot stars likely have less reliable Gaia radial velocity measurements because of their weaker and broader spectral features.

\tabletypesize{\scriptsize}
\movetabledown=1.5in
\begin{rotatetable*}
\begin{deluxetable*}{l c c c c c c c l c c c c}
\tablecaption{Gaia DR3 Compact Object Binary Candidate Sample\label{tab:sample}}

\tablehead{source\_id & RA & Dec & $G$ & P & $M_{1}$ & $M_{\mathrm{2,\,lower}}$ & dv$_{\mathrm{hel}}$/dt & (accel\_ra, accel\_dec) & significance & goodness\_of\_fit & Solution type & References \\ 
 & (deg) & (deg) & (mag) & (d) & (M$_{\odot}$) & (M$_{\odot}$) & (\kms~d$^{-1}$) & (mas~yr$^{-2}$) & & & &}

\startdata
   4373465352415301632\tablenotemark{\footnotesize{a}} &  262.17121 &  \phn$-$0.581092      & 13.77    &  \phn185.8    &  0.95 &  10.5    & & &     13.6 &  \phs0.3 & Orbital & 1,2,3 \\ 
   6328149636482597888\tablenotemark{\footnotesize{b}} &  218.08620 &  $-$10.366347         & 13.34    &  \phn736.0    &  1.09 & \phn2.3  & & &     89.9 &  \phs0.7 & Orbital & 1,4,5,6 \\ 
   5870569352746779008\tablenotemark{\footnotesize{c}} &  207.56972 &  $-$59.239005         & 12.28    &  1352.3       &       &          & & &     39.8 &  \phs3.1 & AstroSpectroSB1 & 1,7,8 \\ 
\hline
   5593444799901901696 &  112.68478 &  $-$30.460563         & 14.42    &  1038.8       &  1.28 & \phn1.8  & & &  \phn6.5 &  \phs0.2  & Orbital & 4 \\ 
   3509370326763016704 &  195.32097 &  $-$18.869145         & 12.47    &  \phn109.4    &  0.68 & \phn3.6  & & &     76.1 &  \phs5.2  & Orbital & 5 \\ 
   3640889032890567040 &  213.17397 &  \phn$-$6.357047      & \phn9.19 &  1076.2       &  1.01 &  79.4    & & &  \phn7.7 &  \phs10.3 & Orbital & 1 \\ 
   6281177228434199296 &  223.20971 &  $-$19.373622         & 11.26    &  \phn153.9    &  0.95 &  10.6    & & &     24.3 &  \phs8.0  & Orbital & 5 \\ 
   6802561484797464832 &  315.10711 &  $-$25.585690         & 12.88    &  \phn574.8    &  1.18 & \phn2.3  & & &  \phn6.8 &  \phs11.0 & Orbital & 5 \\ 
   6588211521163024640 &  329.02330 &  $-$35.377937         & 14.19    &  \phn943.3    &  1.09 & \phn2.0  & & &     10.4 &  \phs4.9  & Orbital & 5 \\ 
   6601396177408279040 &  336.46845 &  $-$32.518802         & 14.07    &  \phn533.5    &  0.99 & \phn2.1  & & &     10.8 &  \phs13.8 & Orbital & 5 \\ 
   4314242838679237120 &  285.68712 &  $+$13.063667         & 17.02    &  1146.0       &  0.62 & \phn1.4  & & &  \phn8.2 &  $-$0.8   & Orbital & 4 \\ 
   6593763230249162112 &  337.21393 &  $-$39.721915         & 13.54    &  \phn679.9    &  1.02 & \phn1.5  & & &     23.3 &  \phs1.7  & Orbital & 4,9 \\ 
\hline
   2000733415898027264 &  333.64320 &  $+$49.973213         & 11.86    &  \phn\phn53.8 &  2.26 &  \phn5.0 & & &  \phn7.1 &  \phs0.0 & SB1 & 1 \\ 
   2086448353089047808 &  298.37120 &  $+$47.813900         & 11.96    &  \phn\phn69.6 &  1.59 &  \phn7.4 & & &     15.9 &   $-$1.8 & SB1 & 1 \\ 
   2226444358294583680 &  339.25503 &  $+$70.542084         & 12.20    &  \phn\phn97.2 &  2.09 &  29.2    & & &  \phn5.9 &  \phs0.6 & SB1 & 1 \\ 
   4060365702574410752 &  263.10178 &  $-$28.733722         & 12.52    &  \phn518.2    &  1.00 &  \phn8.3 & & &     10.2 &  \phs2.1 & SB1 & 1 \\ 
   5625482713303185408 &  132.69420 &  $-$36.226350         & 12.23    &  \phn687.0    &  1.74 &  \phn6.8 & & &  \phn5.8 &   $-$2.1 & SB1 & 1 \\ 
   5846362195472084992 &  214.85468 &  $-$69.780565         & 11.95    &  \phn764.1    &  1.74 &  10.5    & & &  \phn7.0 &  \phs0.2 & SB1 & 1 \\ 
   6059985721200365184 &  189.84145 &  $-$58.698767         & 11.86    &  \phn910.1    &  2.28 &  11.8    & & &  \phn5.5 &   $-$1.2 & SB1 & 1 \\ 
   6102598776102841344 &  216.74790 &  $-$43.515619         & 11.67    &  \phn225.8    &  1.60 &  \phn8.3 & & &     12.9 &   $-$0.1 & SB1 & 1 \\ 
   5352109964757046528 &  159.99924 &  $-$56.402076         & 11.77    &  \phn\phn\phn9.9 &  4.39 & \phn6.6 & & &     13.3 &  \phs0.7 & SB1 & 1 \\ 
\hline
   1864406790238257536 &  309.46725 &  $+$34.891904         & 12.72    &  \phn875.8    &  2.40 &  12.2    & & &     19.6 &  \phs3.3 & AstroSpectroSB1 & 1 \\ 
   3263804373319076480 &  \phn53.73071 & \phn$+$0.152846    & 12.67    &  \phn510.7    &  0.97 &  0.73    & & &     18.1 &  \phs5.6 & AstroSpectroSB1 & 5 \\ 
\hline
   3689209059942075008 &  196.25749 &     \phn$-$0.049722   & 12.64 & & & &   $-$0.026 &  ($-$2.80, 3.23)   &  24.4 &  6.0 & Accel9,Trend2 & 1 \\ 
   1873093722367193216 &  315.54354 &     $+$39.744238      & 12.46 & & & &   $-$0.043 &  ($-$1.07, 0.91)   &  20.8 &  6.3 & Accel9,Trend2 & 1 \\ 
   5358111034809618304 &  156.31622 &     $-$50.452767      & 12.44 & & & &   $-$0.012 &  (0.97, 1.32)      &  49.1 &  9.6 & Accel7,Trend2 & 1 \\ 
    409488405416253696 &  \phn26.28813 &  $+$55.905174      & 12.37 & & & &   $-$0.038 &  ($-$0.94, 0.71)   &  26.7 &  7.9 & Accel7,Trend2 & 1 \\ 
   4169598884257076864 &  262.17483 &     \phn$-$7.081493   & 12.93 & & & &  \phs0.052 & ($-$1.58, $-$1.71) &  34.9 &  6.8 & Accel7,Trend1 & 1 \\ 
   5939087648856671232 &  254.97024 &     $-$47.279485      & 12.41 & & & &   $-$0.184 &  (0.69, 0.54)      &  21.3 &  8.3 & Accel9,Trend1 & 1 \\ 
    392363030772060672 &  \phn\phn3.35335 & $+$47.682088    & 11.39 & & & &  \phs0.011 & ($-$0.23, $-$0.59) &  20.6 &  7.7 & Accel7,Trend2 & 1 \\ 
   2044799672971280384 &  293.02595 &     $+$31.114685      & 12.47 & & & &  \phs0.030 & ($-$0.39, $-$2.61) &  21.3 &  9.6 & Accel9,Trend2 & 1 \\ 
   5309744205505321600 &  139.12199 &     $-$57.341473      & 11.71 & & & &   $-$0.148 &  (0.18, 1.00)      &  30.1 &  6.5 & Accel7,Trend1 & 1 \\ 
   5937774698871600512 &  251.06714 &     $-$49.611861      & 12.37 & & & &   $-$0.062 &  (0.73, 0.65)      &  24.3 &  6.5 & Accel9,Trend2 & 1 \\ 
   5530445257521690624 &  115.80898 &     $-$47.655122      & 12.53 & & & &  \phs0.015 & (2.96, $-$0.37)    &  22.2 &  5.0 & Accel9,Trend1 & 1 \\ 
\enddata
\tablenotetext{a}{Gaia BH1}
\tablenotetext{b}{Gaia NS1}
\tablenotetext{c}{Gaia BH2}
\tablecomments{References: (1) \citet{gaiadr3arenou}, (2)
  \citet{elbadry23a}, (3) \citet{chakrabarti23}, (4)
  \citet{andrews22}, (5) \citet{shahaf23}, (6) \citet{elbadry24a}, (7) 
  \citet{tanikawa23}, (8) \citet{elbadry23b}, (9) \citet{elbadry24b}.  The component masses for Gaia~BH2 are not included in the table; they were not determined in the DR3 orbital catalog because the luminous star is a giant rather than a main sequence star (see the lower right panel of Fig.~\ref{fig:sample_4panel}.}
\end{deluxetable*}
\end{rotatetable*}

\subsection{Observations}

We have carried out a spectroscopic follow-up campaign on the sample
described in \S~\ref{sec:sample}, primarily relying on the MIKE
spectrograph \citep{bernstein03} on the Magellan Clay telescope and
the Levy Spectrograph \citep{vogt14} on the Automated Planet Finder (APF)
Telescope at Lick Observatory.  We used these data to characterize the
luminous component of each binary and obtain precise radial velocity
measurements.  We also obtained a few observations with the IMACS
spectrograph \citep{dressler06,dressler11} on the Magellan Baade
telescope and the FEROS spectrograph \citep{kaufer99} on the 2.2~m ESO/MPG telescope and observed one star with the HIRES spectrograph \citep{vogt94} on the Keck~I telescope.

Our MIKE observations extended from 2022 August through 2026 March.  We
observed with the $0\farcs7$ slit, which provides a spectral
resolution of $R=31000$ with the red camera and $R=40000$ with the
blue camera.  For cool targets (spectral types later than
approximately F5), we aimed to acquire at least one spectrum at
$\mathrm{S/N} \gtrsim 100$~pix$^{-1}$ for stellar parameter
determination, with additional spectra for radial velocity
measurements typically at $\mathrm{S/N} = 30$--50~pix$^{-1}$.  For
hotter, rapidly rotating targets with broad lines (spectral types of A
through early F) we attempted to reach $\mathrm{S/N} > 100$~pix$^{-1}$
for all spectra in order to enable improved estimates of the stellar
continuum and the identification of weak and broad absorption lines.
In some cases, poor observing conditions prevented us from achieving
this S/N goal.

Our APF observations extended from 2022 June through 2024 July.  We observed with a slit width of 2\arcsec\ and a slit length of either 3\arcsec\ or 8\arcsec, providing a spectral resolution of $R=80000$.  Integration times per epoch ranged from 600~s to 2500~s, divided into three frames for longer exposure times.  Because of the smaller aperture and higher spectral resolution of APF, the S/N of these spectra is generally low and we did not attempt to measure anything other than velocities from the APF data.

FEROS observing procedures were described by \citet{elbadry23a,elbadry24b}.

Our IMACS observations consisted of longslit spectra taken with the
$f$/4 camera and the 1200~$\ell$/mm grating blazed at 9000~\AA,
providing $R=11,000$ spectra over a $\sim1400$~\AA\ range covering the
Ca triplet and Paschen series lines and the telluric A-band
absorption.

\subsection{Data Reduction}

We reduced the MIKE data with the Carnegie Python package first
described by \citet{kelson03}.  APF spectra were processed with the
California Planet Search pipeline \citep{rosenthal21}.  We reduced the
IMACS data with the COSMOS data reduction package \citep{oemler17} and
a modified version of the DEEP2 data reduction pipeline
\citep{cooper12,newman13}, following the procedures described in a
number of previous papers \citep[e.g.,][]{simon17,simon20}.

\subsection{Velocity Measurements}

\subsubsection{MIKE}

For MIKE observations of relatively cool stars exhibiting narrow
lines, we measured velocities using only the data from the red
spectrograph ($\lambda \gtrsim 4900$~\AA).  The lack of sky emission lines
at shorter wavelengths may make the wavelength solution less
accurate on the blue side, and the red spectra contain plenty of
absorption lines to enable precise velocities.  We determined
velocities from these data via a $\chi^{2}$ fit to the radial velocity
standard star HD~126053 \citep{stefanik99,soubiran18}, which is a slightly metal-poor G1V star with
similar stellar parameters to most of our targets.  We assumed a heliocentric velocity for HD~126053 of $v_{\rm hel} = -19.21$~\kms\ from Gaia DR3.  We excluded spectral orders
containing significant amounts of telluric absorption and those
covering the Na~D lines (which may be contaminated by interstellar
absorption as well as a forest of telluric features).  We fit each
order independently, with the stellar velocity taken to be the mean of
the $\sim20$ included orders and the uncertainty defined as the
standard deviation of the order by order velocities.  This procedure
is identical to that used for observations of Gaia~BH1 by
\citet{chakrabarti23} and is based on those employed by \citet{sg07} and
\citet{simon17}.

For MIKE observations of hotter stars with rapid rotation
($T_{\mathrm{eff}} \gtrsim 6500$~K), the number of strong absorption
lines in the red data is much lower, so we used both the red and the
blue data for velocities, with most of the RV signal coming from the
blue side for the hottest targets.  As mentioned above,
the wavelength solution for the blue spectrograph could be less
accurate, but since these stars have broad lines, the achievable
velocity precision is lower to begin with, and using the full
wavelength range provides a net benefit to the velocities.  For each
hot target, we obtained a MIKE spectrum of a bright star with similar
effective temperature and rotation velocity to serve as a
template.\footnote{The selected template stars were HD~153232 (F6V; \citealt{pecaut12}) for
  Gaia~DR3~5593444799901901696, HD~10148 (F0V; \citealt{houk88}; $v_{hel} = 13.29$~\kms) for
  Gaia~DR3~5846362195472084992, HD~212728 (A3; \citealt{cp93}; $v_{hel} = -13.77$~\kms) for
  Gaia~DR3~5625482713303185408, HD~218108 (A2; \citealt{cp93}; $v_{hel} = -4.72$~\kms) for Gaia DR3~5939087648856671232 and
  Gaia~DR3~5309744205505321600, and HD~105233 (F1V; \citealt{pecaut12}; $v_{hel} = -14.16$~\kms) for
  Gaia~DR3~4169598884257076864 and Gaia~DR3~5937774698871600512.} We
then carried out a $\chi^{2}$ fit of the target star with the template
spectrum order by order, as described above.  We decided which orders
to include in the velocity measurement for each star based on the
strength of the spectral features in the order, the quality of the
match with the template spectrum, the symmetry of the $\chi^{2}$
minimum as a function of velocity offset, the statistical uncertainty on the velocity for the order, and the agreement with the velocity measured in the orders
with the strongest features.  As for the cooler stars, the stellar
velocity is defined to be the mean of the velocities from the accepted
spectral orders and the uncertainty is the standard deviation of those
velocities.

\subsubsection{APF}

Our approach to measuring velocities from the APF spectra was similar to the procedures used for MIKE.  For cool stars, we used the spectral orders covering wavelengths from 4450--6800~\AA.  At shorter wavelengths the S/N was generally too low, and the orders at longer wavelengths were impacted by telluric absorption, fringing, and sky emission lines.  We used an APF observation of HD~12846 \citep{soubiran18}, also a slightly metal-poor sun-like star, as a template spectrum, assuming a velocity of $v_{\rm hel} = -4.65$~\kms\ from Gaia~DR3.  We fit each target spectrum with the template as described above, defining the stellar velocity to be the mean of the included orders and the uncertainty to be the standard deviation of the order velocities.

For APF observations of hot stars, we used either the A0 star HR~5849 ($v_{\rm hel} = -12.1$~\kms; \citealt{gontcharov06}) or a smoothed version of the HD~12846 spectrum as a template.  Measuring velocities for these stars was generally challenging because of the low S/N of the data and the very broad lines.  In some cases, the only spectral feature that produced a useful velocity was H$\beta$.

\subsubsection{IMACS}

We measured velocities from IMACS spectra using the method described by \citet{simon17}.  The template star for most observations was HD~122563, as in previous IMACS analyses, except for the hottest stars, which used HD~161420 or HD~218108.

\subsubsection{FEROS}

We measured velocities from FEROS spectra as described by \citet{elbadry24a} using synthetic template spectra.  In cases where the formal velocity uncertainty was below 0.1~\kms, we imposed a minimum uncertainty of 0.1~\kms.

\subsubsection{Data availability}

The velocity measurements for all stars are listed in Table~\ref{tab:velocitiess}.  Archival velocities from LAMOST \citep{lamost,luo26} and APOGEE \citep{saydjari25} are included where available.

\begin{deluxetable*}{lcccl}
\tablecolumns{5}
\tabletypesize{\footnotesize}
\tablecaption{\label{tab:velocitiess}Stellar Velocities}
\tablehead{ Gaia source\_id & Julian date & Velocity & Velocity uncertainty & Instrument \\ 
 & & (km~s$^{-1}$) & (km~s$^{-1}$) & }
\startdata
 5593444799901901696 & 2459908.66 &  12.80 &  3.00 &                  FEROS \\
 5593444799901901696 & 2459920.67 &  11.10 &  3.00 &                  FEROS \\
 5593444799901901696 & 2459994.78 &  22.00 &  3.00 &                  FEROS \\
 5593444799901901696 & 2459977.73 &  16.10 &  1.50 &                   MIKE \\
 5593444799901901696 & 2460037.53 &  19.90 &  1.00 &                   MIKE \\
 5593444799901901696 & 2460038.58 &  19.20 &  3.00 &                  FEROS \\
 5593444799901901696 & 2460040.63 &  14.30 &  3.00 &                  FEROS \\
 5593444799901901696 & 2460076.50 &  21.40 &  0.70 &                   MIKE \\
 5593444799901901696 & 2460095.49 &  21.50 &  1.50 &                   MIKE \\
 5593444799901901696 & 2460281.79 &   6.20 &  0.90 &                   MIKE \\
 5593444799901901696 & 2460345.70 &   3.40 &  1.10 &                   MIKE \\
 5593444799901901696 & 2460362.73 &   2.70 &  1.50 &                   MIKE \\
 5593444799901901696 & 2460405.35 &   4.60 &  1.00 &                   MIKE \\
 5593444799901901696 & 2460790.54 &  12.10 &  1.00 &                   MIKE \\
\enddata
\end{deluxetable*}

\subsection{Stellar Parameter Determination}

\subsubsection{Cool stars}
\label{sec:coolstars}

For the program stars that are cool enough for a classical spectroscopic analysis, we derived photospheric and fundamental stellar parameters using the algorithm described by \citet{reggiani2021,reggiani2022} that makes use of both a spectroscopic approach\footnote{The classical spectroscopic approach simultaneously minimizes the line-by-line iron abundance difference between \ion{Fe}{1} and \ion{Fe}{2}-based abundances, as well as their dependencies on excitation potential 
and reduced equivalent widths, to infer T$_{\text{eff}}$, $\log{g}$, and [Fe/H].} and isochrones to infer precise and self-consistent photospheric (T$_{\text{eff}}$, $\log{g}$, and [Fe/H]) and fundamental (mass, luminosity, and radius) stellar parameters.  

For the isochrone fitting we used multiwavelength photometry: Gaia DR3 G \citep{gaia16a,gaia2018,arenou2018,evans2018,hambly2018,riello2018, gaia2021,fabricius2021,lindegren2021a,lindegren2021b,torra2021}, SkyMapper DR4  \textit{u, v, g, r, i, and z} magnitudes \citep{Onken:2024PASA...41...61O},  the Sloan Digital Sky Survey (SDSS) DR16 \citep{SDSS16:2020ApJS..249....3A}, J, H, and Ks bands from the Two Micron All Sky Survey (2MASS) All-Sky Point Source Catalog \citep[PSC,][]{skrutskie2006}, W1 and W2 bands from the Wide-field Infrared Survey Explorer (WISE) AllWISE mid-infrared catalog \citep{wright2010,mainzer2011}, and \textit{g, r, i, z, y} from the PANSTARRS1 (PS1) catalog \citep{PS1:2020ApJS..251....7F}. The included data for each star were based on availability and on data flag-based cuts, as detailed in \cite{Nataf:2024ApJ...976...87N}. We also included the Gaia DR3-based photogeometric distances from \citet{bailer-jones2021} of our targets in our priors.  Finally, we included extinction ($A_V$) inferences from the \textit{Bayestar} model \citep{Green:2019ApJ...887...93G} when available. When \textit{Bayestar} was not available we used extinctions from the SFD maps \citep{Schlegel:1998ApJ...500..525S,SFD:2011ApJ...737..103S}. In both cases, we used the \textit{dustmaps}\footnote{https://dustmaps.readthedocs.io/en/latest/maps.html} code to interpolate extinction values from the corresponding map.

For the spectroscopic-based inferences ([Fe/H] and microturbulent velocities) we used the equivalent widths (EWs) of \ion{Fe}{1} and \ion{Fe}{2} atomic absorption lines. The EWs were measured from our spectra using Gaussian profiles with the \texttt{Spectroscopy Made Harder}\footnote{https://github.com/andycasey/smhr} semi-automated code. All EWs were individually reviewed, and bad fits were removed from the analysis. We assumed \citet{asplund2021} photospheric solar abundances. 

As described in detail in \citet{reggiani2022}, we used the \texttt{isochrones} package\footnote{\url{https://github.com/timothydmorton/isochrones}} \citep{morton2015} to fit the MESA Isochrones and Stellar Tracks \cite[MIST;][]{dotter2016,choi2016,paxton2011,paxton2013,paxton2015,paxton2018,paxton2019} library to our photospheric stellar parameters as well as our input multiwavelength photometry, parallax, and extinction data using \texttt{MultiNest}\footnote{\url{https://ccpforge.cse.rl.ac.uk/gf/project/multinest/}} \citep{feroz2008,feroz2009,feroz2019} via \texttt{PyMultinest} \citep{buchner2014}.

Our adopted stellar parameters ($\rm{T_{eff}}$ and surface gravity from the isochrone analysis, and [Fe/H] inferred from the atomic \ion{Fe}{1} and \ion{Fe}{2} lines) are listed in Table \ref{table:stellar_params}. The uncertainties from the isochrone analysis, listed in Table \ref{table:stellar_params}, include statistical uncertainties only. That is, they are uncertainties derived under the unlikely assumption that the MIST isochrone grid we use in our analyses perfectly reproduces all stellar properties.  See \citet{reggiani2021,reggiani2022} for some discussion of the systematic uncertainties associated with this approach.

\begin{deluxetable*}{llccc}
\tablecaption{Stellar Parameters and Primary Masses \label{table:stellar_params}}
\tablewidth{0pt}
\tablehead{
\colhead{Star} & \colhead{T$_{\mathrm{eff}}$} & \colhead{$\log{g}$} & \colhead{[Fe/H]} & \colhead{Mass} \\ 
\colhead{} & \colhead{(K)} & \colhead{} & (dex) & \colhead{($M_{\odot}$)}} 
\startdata
\sidehead{\textit{Cool stars}}
Gaia DR3 3509370326763016704 & $4661^{+122}_{-101}$ & $4.66^{+0.02}_{-0.01}$ & \phs$0.08\pm0.15$ & $0.65^{+0.02}_{-0.02}$ \\ 
Gaia DR3 3640889032890567040 & $5969^{+142}_{-115}$ & $4.38^{+0.04}_{-0.04}$ & $-0.29\pm0.03$ & $0.85^{+0.07}_{-0.05}$ \\ 
Gaia DR3 6281177228434199296 & $6261^{+84}_{-108}$ & $4.09^{+0.02}_{-0.02}$ & $-0.30\pm0.09$ & $0.80^{+0.02}_{-0.01}$ \\ 
Gaia DR3 6802561484797464832 & $5970^{+100}_{-98}$ & $4.20^{+0.04}_{-0.04}$ & $-0.17\pm0.05$ & $0.99^{+0.06}_{-0.06}$ \\ 
Gaia DR3 6588211521163024640 & $6013^{+89}_{-70}$ & $4.42^{+0.03}_{-0.04}$ & $-0.20\pm0.04$ & $0.88^{+0.04}_{-0.05}$ \\ 
Gaia DR3 6601396177408279040 & $5847^{+128}_{-122}$ & $4.43^{+0.03}_{-0.04}$ & $-0.41\pm0.13$ & $0.85^{+0.05}_{-0.05}$ \\ 
Gaia DR3 4314242838679237120 & $3724^{+40}_{-26}$ & $4.85^{+0.02}_{-0.02}$ & $-0.14\pm0.20$ & $0.44^{+0.02}_{-0.02}$ \\ 
Gaia DR3 6593763230249162112 & $5783^{+16}_{-14}$ & $4.46^{+0.01}_{-0.02}$ & \phs$0.07\pm0.03$ & $0.99^{+0.01}_{-0.01}$ \\ 
Gaia DR3 3263804373319076480 & $5613^{+46}_{-42}$ & $4.48^{+0.03}_{-0.03}$ & $-0.27\pm0.03$ & $0.77^{+0.02}_{-0.02}$ \\ 
Gaia DR3 3689209059942075008 & $6098^{+112}_{-123}$ & $3.91^{+0.06}_{-0.06}$ & $-0.42\pm0.12$ & $0.94^{+0.08}_{-0.09}$ \\ 
Gaia DR3 5530445257521690624 & $6035^{+115}_{-111}$ & $4.26^{+0.04}_{-0.04}$ & $-0.47\pm0.15$ & $0.98^{+0.06}_{-0.06}$ \\ 
\hline
\sidehead{\textit{Hot stars}}
Gaia DR3 5593444799901901696 &  &  &  & \phn$1.08 \pm 0.04$ \\
Gaia DR3 5625482713303185408 &  &  &  & \phn$1.86 \pm 0.11$ \\
Gaia DR3 5846362195472084992 &  &  &  & $1.64^{+0.05}_{-0.06}$ \\
Gaia DR3 6059985721200365184 &  &  &  & $2.50^{+0.12}_{-0.16}$ \\
Gaia DR3 4169598884257076864 &  &  &  & $1.71^{+0.07}_{-0.10}$ \\
Gaia DR3 5939087648856671232 &  &  &  & $1.41^{+0.10}_{-0.04}$ \\
Gaia DR3 5309744205505321600 &  &  &  & $2.26^{+0.22}_{-0.16}$ \\
Gaia DR3 5937774698871600512 &  &  &  & \phn$1.64 \pm 0.06$ 
\enddata
\tablecomments{This table includes parameters for stars for which we obtained a high enough quality spectrum to determine robust parameters and where the spectrum is dominated by a single stellar component.}
\end{deluxetable*}

\subsubsection{Masses for hot stars}

For stars with temperatures and rotation velocities that are not suitable for EW analysis, we used only the isochrone fitting portion of the analysis from \S~\ref{sec:coolstars} to determine temperatures, surface gravities, and stellar masses.  In these cases, we restricted the metallicity to a range of $-0.2 \le \mbox{[Fe/H]} \le 0.2$; without this constraint, the metallicity tended to run away to unphysically low values for young massive stars.  As with the cooler stars, we used \textit{Bayestar} extinction values when possible, SFD values when not, and estimates from the Planck survey if neither of the others were available. For the low-latitude stars where coarse extinction maps may not be reliable, the estimated extinction values are unrealistically high.  For example, based on the SFD map, extinction for Gaia~DR3~5939087648856671232 is A$_{\textrm{V}}=7.01\pm0.16$~mag, which is unlikely for a star located less than 1~kpc away \citep{bailer-jones2021}. In those cases, we removed the A$_{\textrm{V}}$ information both from our priors and the likelihood, but kept the distance information to inform the Bayesian fitting process. Our final results are physical, and with extinction values that are more plausible.

\subsection{Modeling Gaia Astrometric Orbits Without Epoch Astrometry}

Although Gaia~DR3 provided astrometric orbital solutions for binaries,
it did not include the individual astrometric measurements to which
the orbital fits were made.  This situation has encouraged the
community to explore the additional information available in other
Gaia parameters that can constrain binarity and various orbital
properties.  In particular, the Reduced Unit Weight Error (RUWE;
\citealt{lindegren2021b,halbwachs23}) is elevated for stars exhibiting
astrometric orbital motion \citep[e.g.,][]{belokurov20,penoyre20}.  As
shown by \citet{penoyre20}, orbits with smaller eccentricities and
smaller inclinations result in larger RUWE if the other orbital
properties are held fixed.  The gaiamock forward modeling framework
introduced by \citet{elbadry_forwardmodel} and the companion study by
\citet{lam25} provide convenient implementations for calculating the
RUWE of a binary as it would be measured by Gaia and determining
whether a system would be included in any of the DR3 non-single star
catalogs.

\section{Results}
\label{sec:results}

Although only a handful of radial velocities are necessary to falsify
the Gaia orbital solutions \citep[e.g.,][]{elbadry23a}, we continued
to monitor the orbits of as many targets as possible in order to try
to determine the actual nature of these binary systems.  Once
sufficient velocity measurements had been obtained, we used TheJoker
\citep{joker}, a rejection sampling algorithm, to identify approximate
orbital solutions.  After identifying the likely period, amplitude, and eccentricity with TheJoker, we ran MCMC fits using the emcee package \citep{fm13} to determine the full orbital parameters.

Below we discuss our follow-up of each candidate binary system.

\subsection{Candidates with Astrometric Binary Orbits}

\begin{itemize}

\item{Gaia~DR3~5593444799901901696 was identified by \citet{andrews22}
  as a binary consisting of a $1.27 \pm 0.2$~\msun\ primary and a
  $2.57^{+0.86}_{-0.69}$~\msun\ secondary, making it a candidate for
  containing a neutron star or a mass-gap black hole.  The Gaia~DR3
  astrometric orbital solution for this system has a period of $1039
  \pm 146$~d and a moderate eccentricity of $e=0.44\pm0.07$.  Our
  spectroscopy showed that the primary is a rapidly rotating mid-F
  star\footnote{\citet{elbadry24b} noted that the high rotation velocity prevented them from obtaining accurate velocity measurements with their lower S/N FEROS spectra.} with an estimated mass of $1.4$~\msun.  Using 9 RV epochs from
  MIKE and 5 from FEROS, we determined a period of $20.023^{+0.022}_{-0.025}$~d and a velocity semi-amplitude
  of $K_{1} = 9.4^{+0.6}_{-0.5}$~\kms\ (see Fig.~\ref{fig:5593rvs}).  The resulting
  mass function is $f=0.0016^{+0.0004}_{-0.0003}$~\msun, indicating a minimum companion
  mass of 0.13~\msun.  The companion is therefore likely an M dwarf or
  possibly a white dwarf.  The large discrepancy between the astrometric and RV
  orbits could suggest that this system is in fact a hierarchical
  triple, where the velocities are tracing an inner binary and the
  Gaia orbit is related to an outer tertiary.  (Given the sparse
  sampling of the Gaia astrometry, DR3 data are not sensitive to
  astrometric orbits with periods shorter than $\sim100$~d;
  \citealt{gaiadr3arenou}.)  The RUWE for the source in the DR3 catalog is consistent with the value expected for the Gaia orbital solution, and much larger than would be caused by the spectroscopic binary orbit.  However, our velocity data span 882~d,
  covering most of the Gaia orbital period, and the predicted
  semi-amplitude is $K_{1} = 28$~\kms.  The RV data do not give any
  indication of such large amplitude variations on a long time scale.
  Gaia~DR4 data will be informative for evaluating the possibility of
  a triple system.}

\begin{figure}
    \centering
    \includegraphics[width=0.47\textwidth]{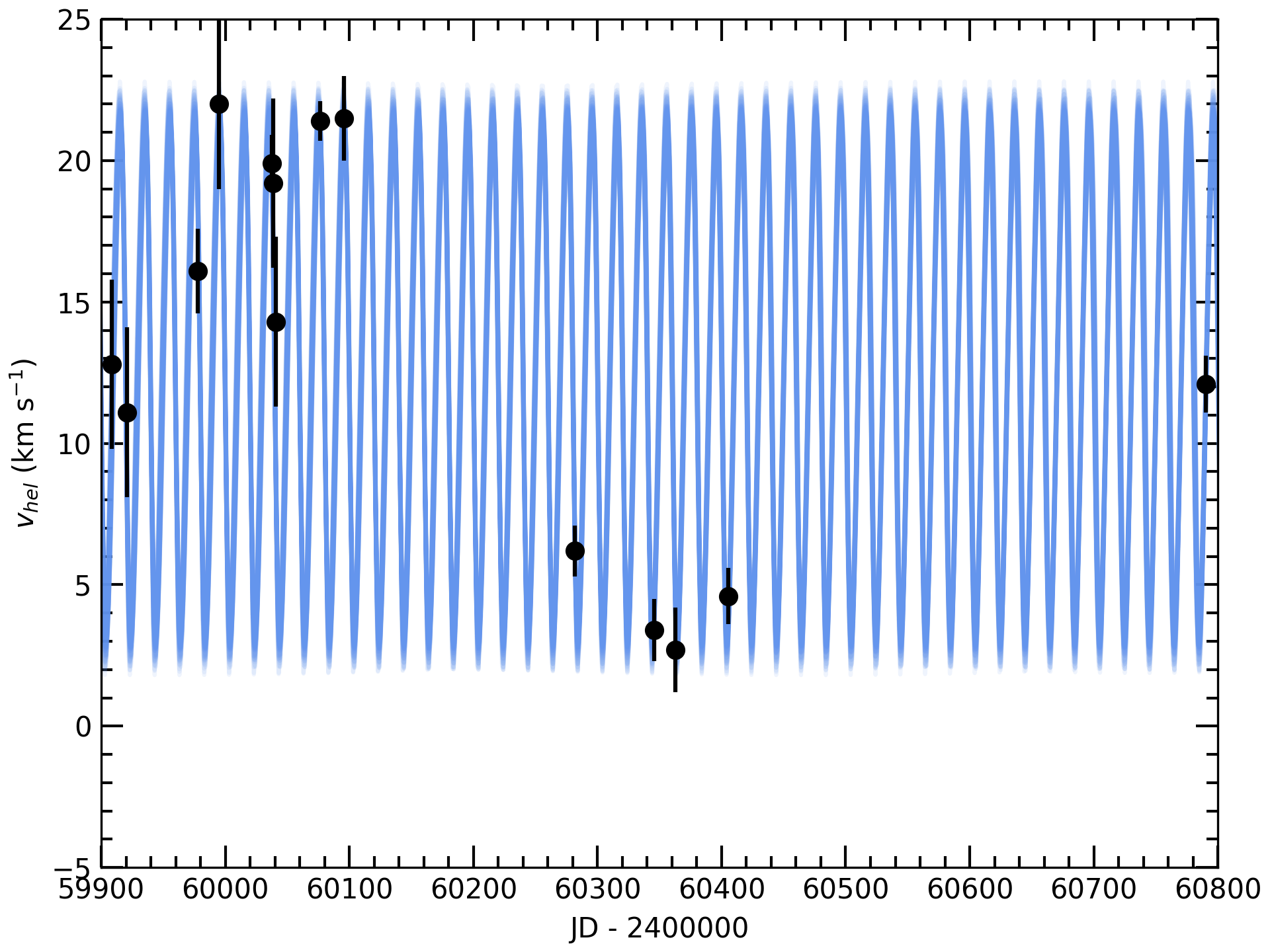}
    \caption{MCMC fit to the radial velocity curve of
      Gaia~DR3~5593444799901901696.  The black points are the RV
      measurements from MIKE and the blue curves are the 100
      best-fitting solutions from the MCMC.  The orbit has a period of
      20~d, a semi-amplitude of 9.4~\kms, and an eccentricity of
      0.14.}\label{fig:5593rvs}
\end{figure}

\item{Gaia~DR3~3509370326763016704 was classified as an AMRF Class~III
  binary by \citet{shahaf23}.  Its DR3 orbital period is $109.39\pm0.06$~d, and
  with an assumed primary mass of $0.68\pm0.05$~\msun, the secondary mass is
  3.6~\msun.  We find a consistent primary mass
  ($M_{1} = 0.65 \pm 0.02$~\msun) but the velocity variability is dramatically lower than
  predicted.  We derived a period of $457.88^{+9.63}_{-6.46}$~d and a semi-amplitude of
  $K_{1} = 1.9^{+2,0}_{-0.3}$~\kms\ (see Fig.~\ref{fig:3509rvs}).  The minimum mass of the companion is therefore
  $\sim0.05$~\msun, indicating a likely brown dwarf unless the orbit
  is rather face-on.  From Kepler's third law, the semi-major axis of
  the orbit is slightly more than 1~AU.  Using the equations in
  \citet{lam25}, the expected size of the orbit photocenter for this
  system is $\sim0.35$~mas, which is small but should be detectable in
  Gaia~DR3.  Applying the forward modeling framework from
  \citet{lam25} to this source, we find that such a binary should have
  a $\gtrsim50$\% probability of being correctly included in the DR3
  binary catalog for any inclination angle.  The gaiamock package \citep{elbadry_forwardmodel} predicts a somewhat lower likelihood of 38\% that the orbital solution would have been included in DR3.  The large values of the
  astrometric excess noise (1.0~mas) and RUWE (10.2) for this source
  suggest that something beyond a brown dwarf orbiting a low-mass star
  may be present.  Supporting this possibility is the
  ipd\_frac\_multi\_peak value of 2\%, which is consistent with the
  presence of an almost-resolved luminous companion.  If so, the
  presence of a wide tertiary component may have biased the
  astrometry through blending with the primary.}

\begin{figure}
    \centering
    \includegraphics[width=0.47\textwidth]{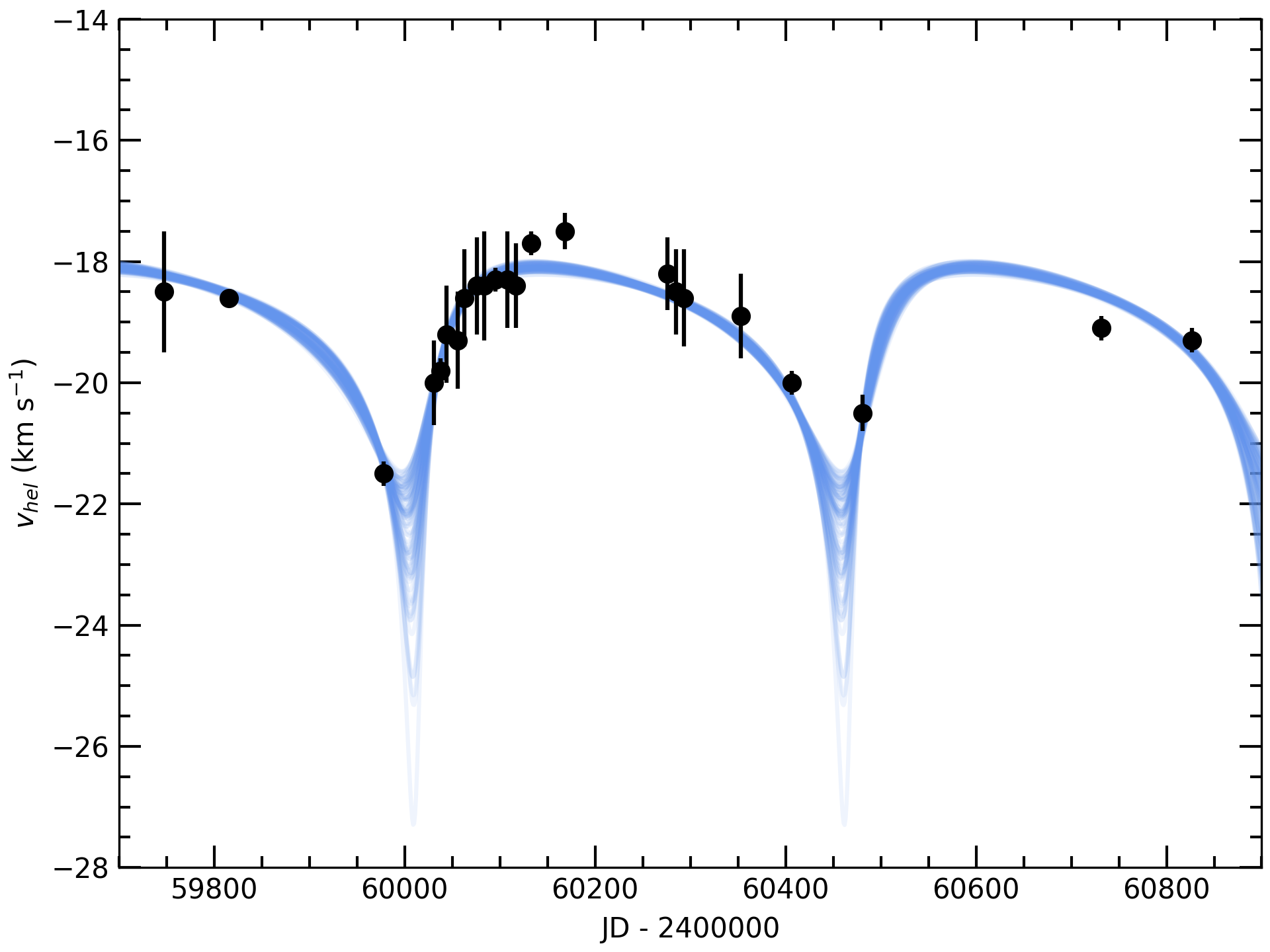}
    \caption{MCMC fit to the radial velocity curve of
      Gaia~DR3~3509370326763016704.  The black points are the RV
      measurements from MIKE and APF and the blue curves are the 100
      best-fitting solutions from the MCMC.  The orbit has a period of
      458~d, a semi-amplitude of 1.9~\kms, and an eccentricity of
      0.54.}\label{fig:3509rvs}
\end{figure}

\item{Gaia~DR3~3640889032890567040 (HD~124273) was listed in the orbital solution
  catalog with a $1076\pm12$~d orbital period and a minimum secondary
  mass of 79.4~\msun.  It was not included in the \citet{andrews22} or
  \citet{shahaf23} catalogs of compact object candidates;
  \citeauthor{andrews22} discussed and dismissed it because of a large
  goodness-of-fit value (although still within the range considered by
  \citealt{halbwachs23} to indicate reasonable fits) and large
  uncertainties on two of the Thiele-Innes parameters.  Nevertheless,
  the unusually high secondary mass would make this an exceptionally
  interesting system if real, and the implied RV signal is very large
  and easy to test.  We verified that this star is indeed a binary, but
  the derived properties bear no resemblance to those from the astrometric orbit. We determined a period of $781.0^{+8.7}_{-7.7}$~d and a semi-amplitude of $K_{1} = 8.2\pm0.3$~\kms (see Fig.~\ref{fig:3640rvs}).  This orbit is consistent with the measured value of rv\_amplitude\_robust, whereas the DR3 orbit predicts a much larger amplitude.  The mass function is $f=0.037^{+0.004}_{-0.003}$~\msun, corresponding to a minimum secondary mass of 0.38~\msun\ and suggesting an M dwarf or white dwarf companion.  Using gaiamock \citep{elbadry_forwardmodel}, we found that this orbit has a 95\% probability of being included in the DR3 binary catalog despite a possible period alias near $P=2$~yr.  We also calculated the RUWE for the RV orbital solution using gaiamock, which showed that the observed RUWE is significantly larger than expected.  It is unclear what is responsible for the large astrometric perturbations seen by Gaia; perhaps an additional luminous companion is present in the system. }

\begin{figure}
    \centering
    \includegraphics[width=0.47\textwidth]{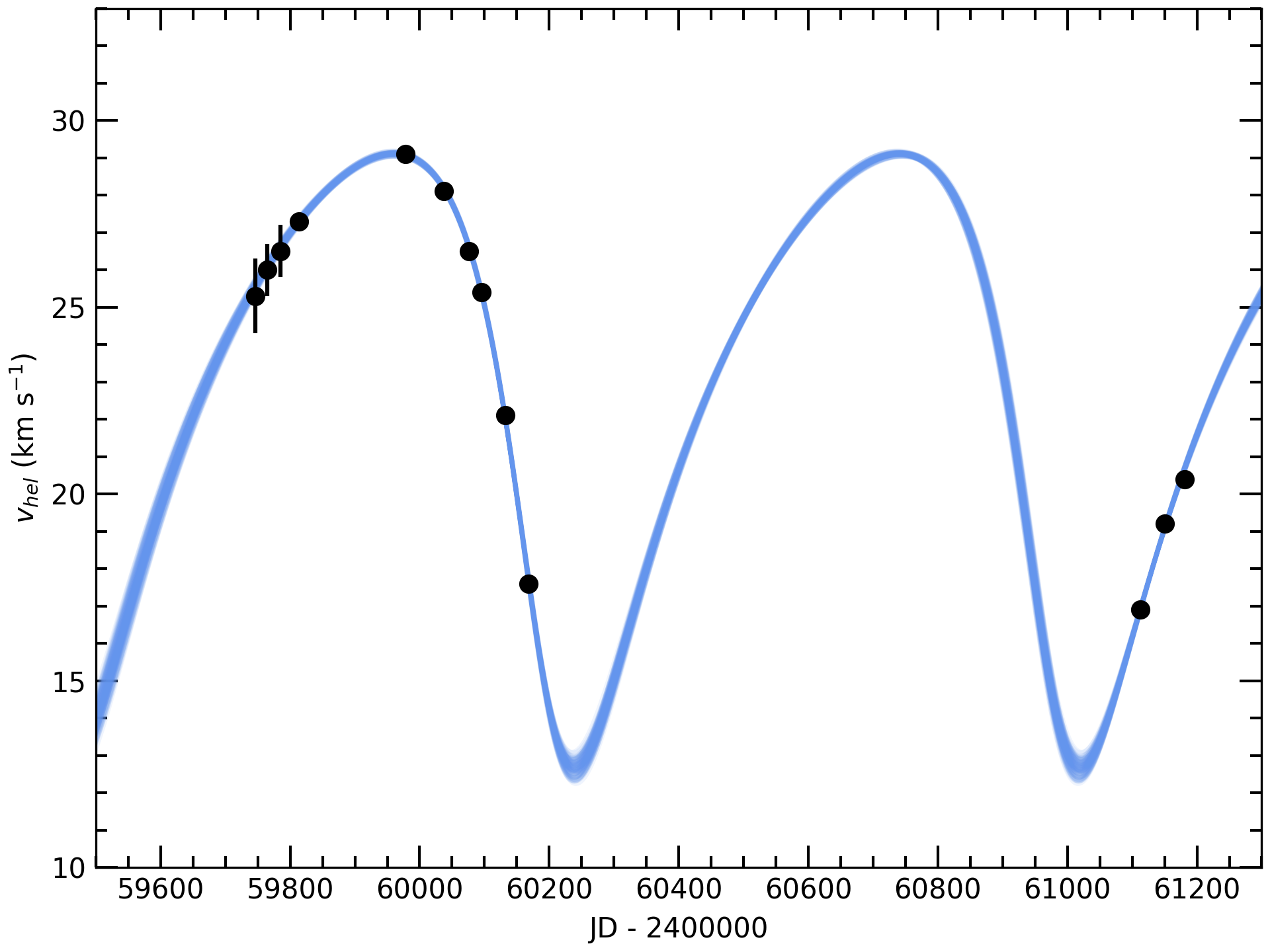}
    \caption{MCMC fit to the radial velocity curve of
      Gaia~DR3~3640889032890567040.  The black points are the RV
      measurements from MIKE and APF and the blue curves are the 100 best-fitting solutions from the MCMC.  The orbit has a period of 781~d, a semi-amplitude of 8.2~\kms, and an eccentricity of 0.34.}\label{fig:3640rvs}
\end{figure}

\item{Gaia~DR3~6281177228434199296 was categorized as an
  AMRF~Class~III system by \citet{shahaf23}.  Its reported properties
  from the astrometric orbit are very similar to those of Gaia~BH1,
  with a period of $153.9 \pm 0.4$~d and a secondary mass of $11.9 \pm 1.3$~\msun consistent
  with a black hole.  The significance of the orbital fit is high
  (24.3), but the goodness of fit value of 8.0 is substantially larger
  than that of BH1 and many other good astrometric solutions.  We
  measured a period of $273.40^{+0.19}_{-0.18}$~d and a semi-amplitude of $K_{1} =
  15.94^{+0.12}_{-0.11}$~\kms\ (see Fig.~\ref{fig:6281rvs}), indicating a secondary of
  at least 0.48~\msun.  However, we also detected a weak second set of
  absorption lines in our spectra, making this system an SB2 binary
  containing two luminous stars.  The detection of a fainter secondary
  with narrow stellar lines constrains the secondary mass to be lower than the primary mass of 
  0.8~\msun, and therefore the inclination of the binary
  to be $>37\degr$.
  Forward modeling indicates that this binary system should have had a
  reasonable ($p \approx 0.6$) of being correctly detected and measured
  in DR3.  We also note that the radial velocity amplitude of this
  star measured by Gaia is 20.1~\kms, roughly consistent with the
  binary amplitude we determined (although perhaps biased by the
  presence of the secondary star), and less likely to be consistent
  with a black hole companion.}

\begin{figure}
    \centering
    \includegraphics[width=0.47\textwidth]{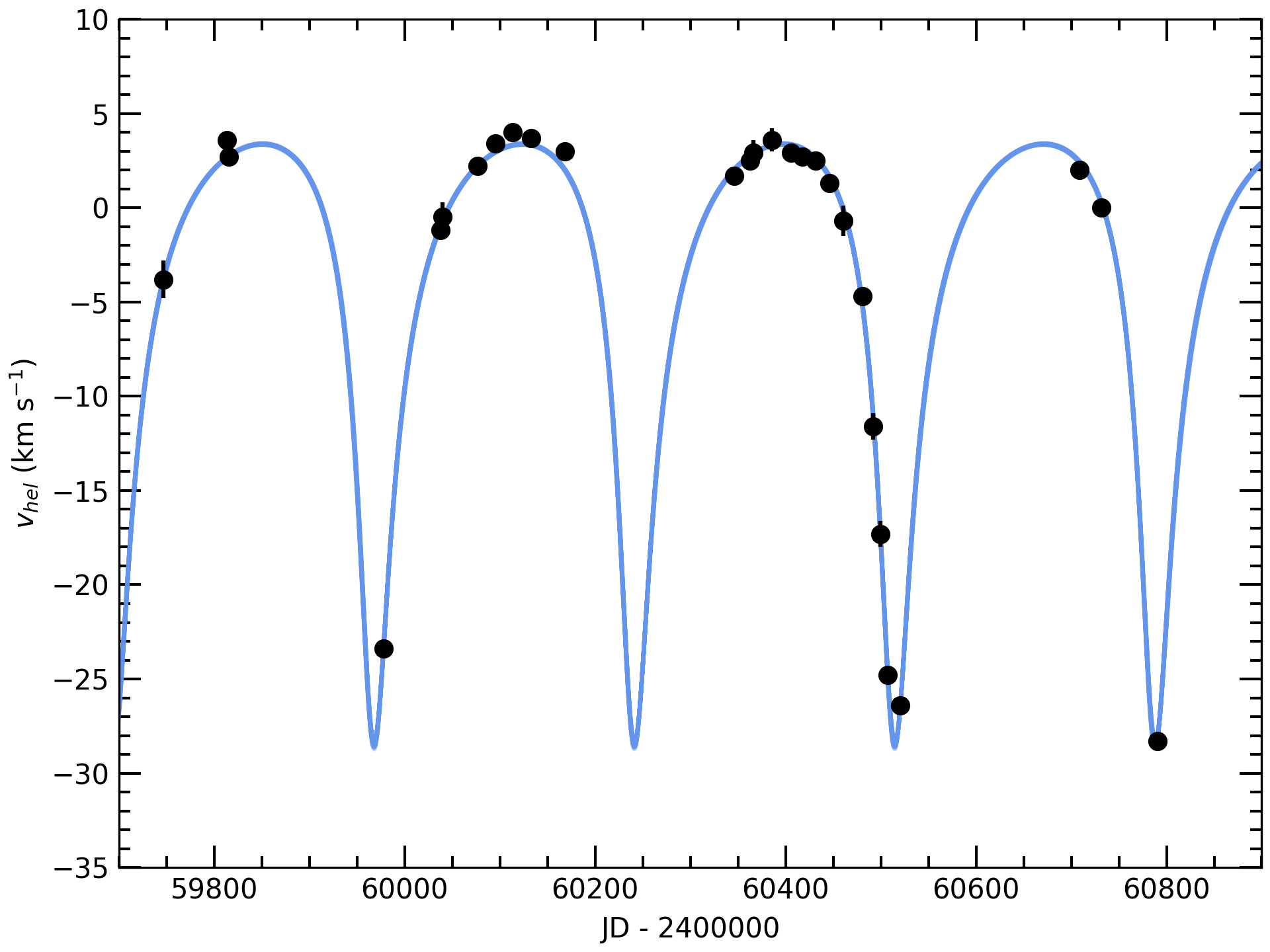}
    \caption{MCMC fit to the radial velocity curve of
      Gaia~DR3~6281177228434199296.  The black points are the RV
      measurements from MIKE, APF, and FEROS and the blue curves are the 100
      best-fitting solutions from the MCMC.  The orbit has a period of
      273~d, a semi-amplitude of 15.9~\kms, and an eccentricity of
      0.55.}\label{fig:6281rvs}
\end{figure}

\item{Gaia~DR3~6802561484797464832 was also an AMRF~Class~III system
  from \citet{shahaf23}, with a period of $575 \pm 6$~d, a large eccentricity
  of $0.83\pm0.07$, and a secondary mass of $3.1^{+1.0}_{-0.8}$~\msun\ pointing to a candidate
  black hole.  From our velocity measurements, we derived a slightly
  shorter period of $524.8^{+5.4}_{-5.6}$~d and a small semi-amplitude of $K_{1} =
  2.6^{+0.8}_{-0.2}$~\kms\ (see Fig.~\ref{fig:6802rvs}).  The mass function is
  therefore 0.00073~\msun\ and the minimum secondary mass is
  0.10~\msun.  As with Gaia~DR3~3640889032890567040 and
  Gaia~DR3~6281177228434199296, the goodness of fit for this star is
  high, and the significance is also relatively low (6.8), perhaps
  contributing to the incorrect astrometric fit.  The \citet{lam25}
  forward modeling suggests a low probability of this binary being
  included in the DR3 catalog unless it is observed face-on ($i
  \lesssim 10\degr$); the gaiamock probability is also relatively low (20\%).  The DR3 RUWE value is substantially lower than should be the case for the Gaia orbital solution, but larger than implied by the spectroscopic orbit, suggesting that the inclination angle may indeed be low.  Nevertheless, the mass function is so small that the companion would remain in the M dwarf or white dwarf mass range unless $i < 4\degr$.}

\begin{figure}
    \centering
    \includegraphics[width=0.47\textwidth]{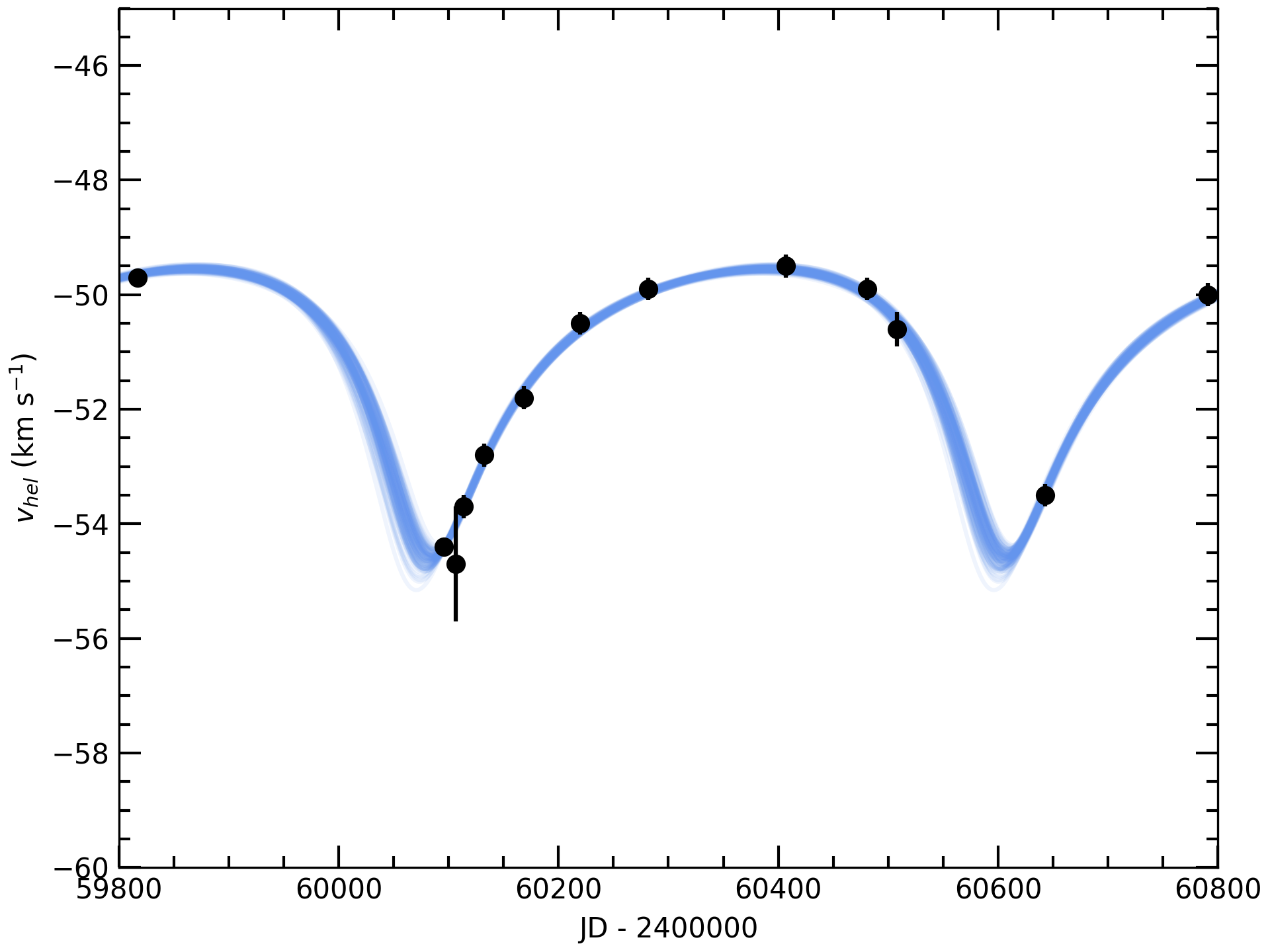}
    \caption{MCMC fit to the radial velocity curve of
      Gaia~DR3~6802561484797464832.  The black points are the RV
      measurements from MIKE and the blue curves are the 100
      best-fitting solutions from the MCMC.  The orbit has a period of
      525~d, a semi-amplitude of 2.6~\kms, and an eccentricity of
      0.46.}\label{fig:6802rvs}
\end{figure}

\item{Gaia~DR3~6588211521163024640 was another AMRF~Class~III binary
  from \citet{shahaf23}.  Its DR3 period is $943 \pm 45$~d, with a very
  eccentric ($e = 0.97 \pm 0.12$) orbit and a secondary mass of $2.4^{+0.5}_{-0.4}$~\msun,
  potentially placing it in the mass gap between neutron stars and black holes.
  We derived a radial velocity period of $641.4^{+4.6}_{-4.1}$~d and a semi-amplitude of
  $K_{1} = 4.84^{+0.05}_{-0.04}$~\kms\ (see Fig.~\ref{fig:6588rvs}).  The secondary
  thus has a minimum mass of 0.19~\msun, likely corresponding to an M
  dwarf.  The goodness of fit value for the DR3 solution is 4.9, lower
  than those of most of the spurious fits discussed above.  A binary
  with the derived parameters would need an inclination $i \lesssim
  20\degr$ to have a high probability ($p > 0.5$) of appearing in the
  DR3 catalog, on average the detection probability is around 30\%.  Forward modeling with the gaiamock or \citet{lam25} frameworks shows that the RUWE is consistent with the spectroscopic orbit and smaller than it should be for the Gaia orbit.  The inclination angle required in order to reproduce the observed RUWE value is $\sim20$--50\degr}, suggesting an actual secondary mass of 0.25--0.6~\msun.

\begin{figure}
    \centering
    \includegraphics[width=0.47\textwidth]{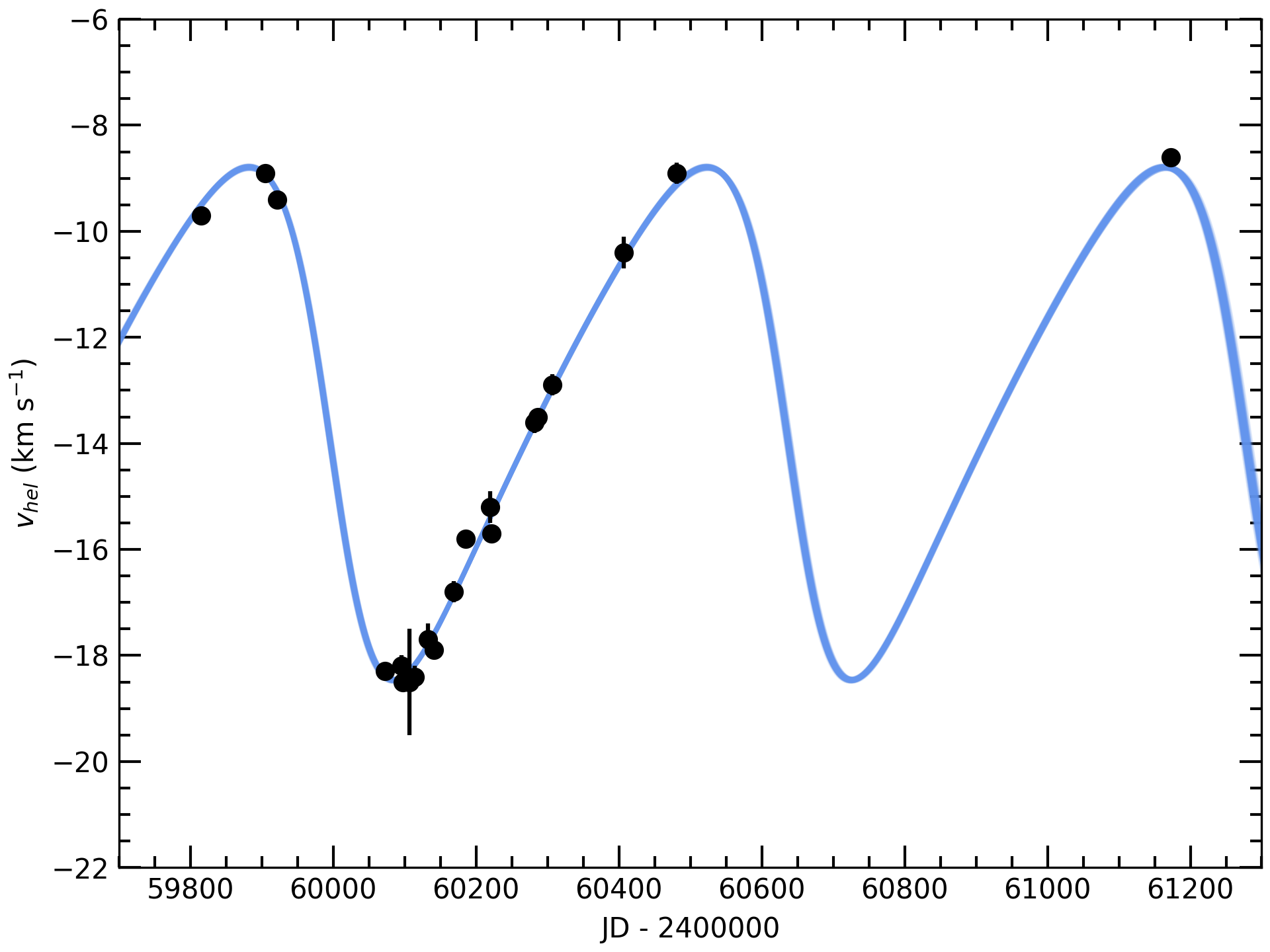}
    \caption{MCMC fit to the radial velocity curve of
      Gaia~DR3~6588211521163024640.  The black points are the RV
      measurements from MIKE and the blue curves are the 100
      best-fitting solutions from the MCMC.  The orbit has a period of
      641~d, a semi-amplitude of 4.8~\kms, and an eccentricity of
      0.30.}\label{fig:6588rvs}
\end{figure}

\item{Gaia~DR3~6601396177408279040 was the final AMRF~Class~III source
  \citet{shahaf23} identified in the DR3 orbital catalog.  It has a
  period of $533.5 \pm 2.0$~d and a secondary mass of $2.6^{+0.5}_{-0.4}$~\msun.  The RV period of
  $525.5^{+8.9}_{-7.9}$~d agrees with the astrometric period within the uncertainties,
  but the semi-amplitude is only $K_{1} = 1.46^{+0.10}_{-0.09}$~\kms\ rather than the
  predicted 51~\kms\ (see Fig.~\ref{fig:6601rvs}).  As for several
  other sources, this result identifies the secondary as a likely
  brown dwarf or possibly a very low-mass star ($M\sin{i} = 0.05$~\msun).  Similar to those
  cases, the goodness of fit for the DR3 solution is high
  (13.8).  Given the long period and very low velocity amplitude, the probability of the orbit being included in DR3 is $\sim50$\% unless the system is relatively face-on.  However, the RUWE is much higher than should be produced by a binary following the spectroscopic orbit.}

\begin{figure}
    \centering
    \includegraphics[width=0.47\textwidth]{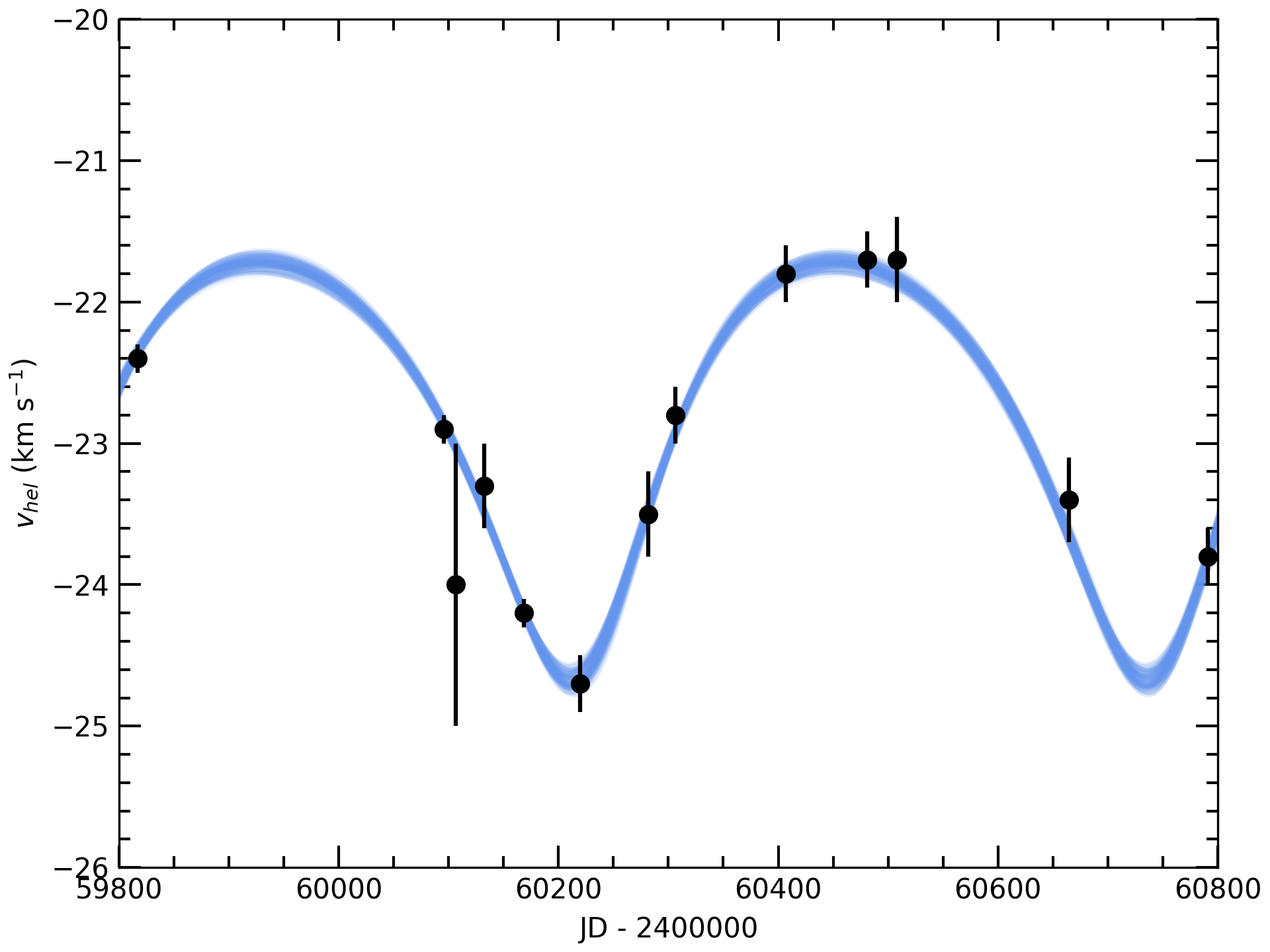}
    \caption{MCMC fit to the radial velocity curve of
      Gaia~DR3~6601396177408279040.  The black points are the RV
      measurements from MIKE and the blue curves are the 100
      best-fitting solutions from the MCMC.  The orbit has a period of
      526~d, a semi-amplitude of 1.5~\kms, and an eccentricity of
      0.26.}\label{fig:6601rvs}
\end{figure}

\item{Gaia~DR3~4314242838679237120 was classified by \citet{andrews22}
  as a binary system with a $<1$~\msun\ primary and a
  2.25~\msun\ secondary on an eccentric, long-period ($1146\pm191$~d) orbit,
  making it a candidate wide neutron star binary.  It is the faintest
  system ($G=17.02$) identified in DR3 with a candidate dark companion
  above 2~\msun, more than 2~mag fainter than any of our other
  targets.  It is too faint to be observed with APF, but we obtained
  three low S/N MIKE spectra to measure its velocity.  Over 436~d, the 
  velocity remained constant within $1.4 \pm 0.9$~\kms.  A $\chi^{2}$ 
  test of the hypothesis that the velocity is constant returns a p-value
  of 0.31, providing no evidence for orbital motion.  This star has no 
  Gaia velocity data because of its magnitude, but the MIKE measurements
  are sufficient to rule out the Gaia orbital solution (see Fig.~\ref{fig:4314rvs}).  This star does
  not appear to be a binary, although a low-mass companion or a long orbital period cannot be excluded from the RV measurements.}

\begin{figure}
    \centering
    \includegraphics[width=0.47\textwidth]{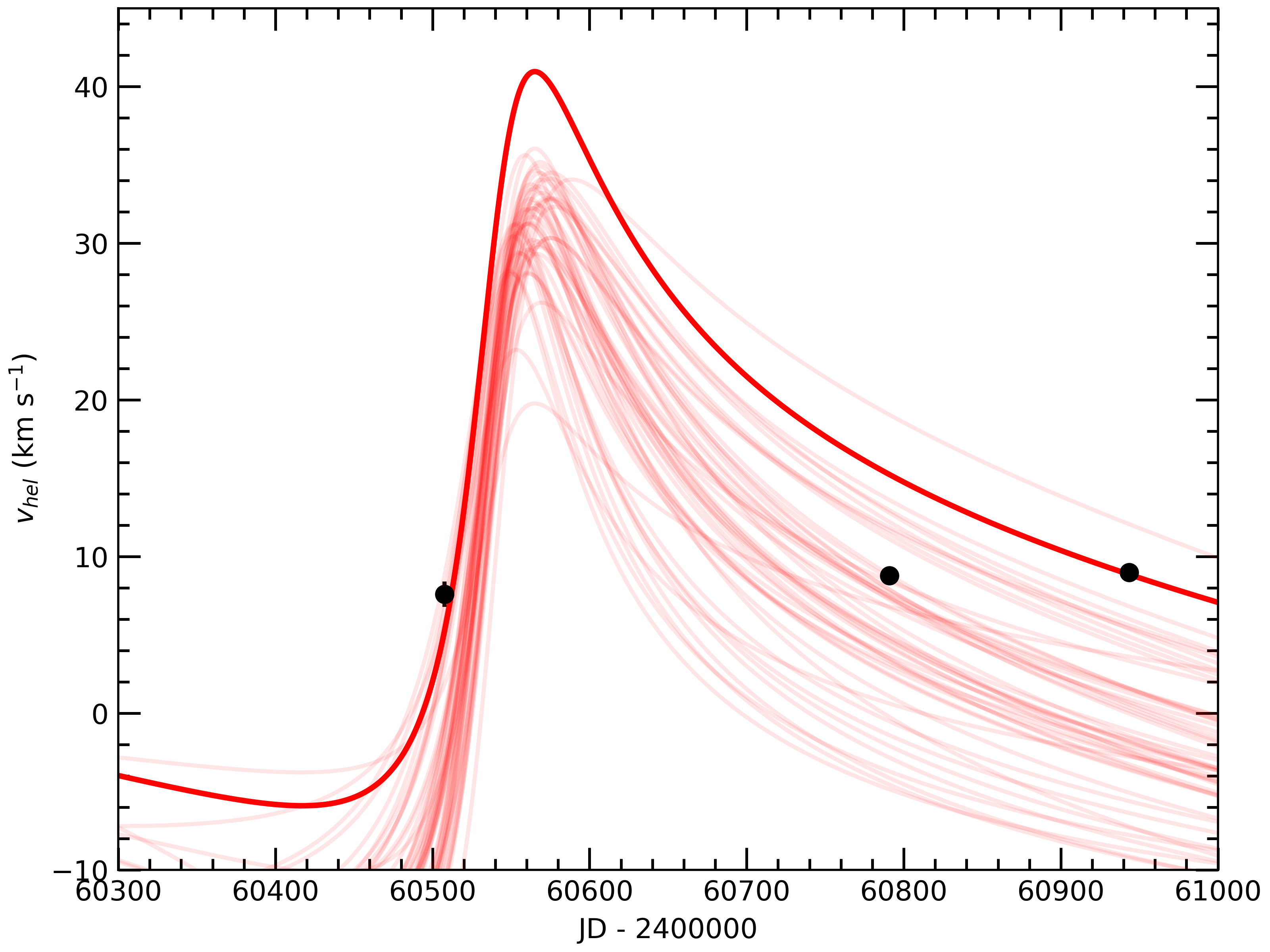}
    \caption{Predicted radial velocity curve (red) based on the Gaia DR3 astrometric orbit for 
      Gaia~DR3~4314242838679237120 compared to observed velocities (black).  The center of mass velocity for the system is not known, so to compute the prediction from the DR3 orbital solution we assumed the mean velocity from the 3 MIKE measurements.  To indicate the uncertainty on the predicted RV curve, we constructed an additional 50 possible RV curves (partially transparent red lines) by drawing random samples of the Thiele-Innes elements and their uncertainties.  Although some of the RV curves agree with two of the three data points, none are consistent with all three.  We therefore rule out the Gaia orbit.}\label{fig:4314rvs}
\end{figure}

\item{Gaia~DR3~6593763230249162112 was selected by \citet{andrews22}
  and \citet{shahaf23} as a compact object binary candidate with a
  period of $680\pm3$~d and a secondary mass of $1.69 \pm 0.16$~\msun.
  \citet{elbadry24b} obtained five RV measurements and ruled out the
  DR3 orbital solution.  However, since their RVs appeared plausibly
  consistent with the Gaia orbit with a significant phase shift, we
  decided to obtain additional data.  We obtained five new velocities
  with MIKE, yielding a period of $951.5^{+6.3}_{-7.5}$~d when combined with the
  \citet{elbadry24b} data (see Fig.~\ref{fig:6593rvs}).  The semiamplitude of the binary is $7.3 \pm 0.2$~\kms,
  and the minimum companion mass is 0.42~\msun.  The companion is most
  likely an M dwarf or a white dwarf unless the orbit is quite face
  on.  The RUWE value is consistent with the derived orbit and a factor of $\sim2$ smaller than would be caused by the Gaia orbit.  The inclination angle implied by the RUWE value is $\sim60\degr$, which would yield a secondary mass of $\sim0.50$~\msun.}
  
\begin{figure}
    \centering
    \includegraphics[width=0.47\textwidth]{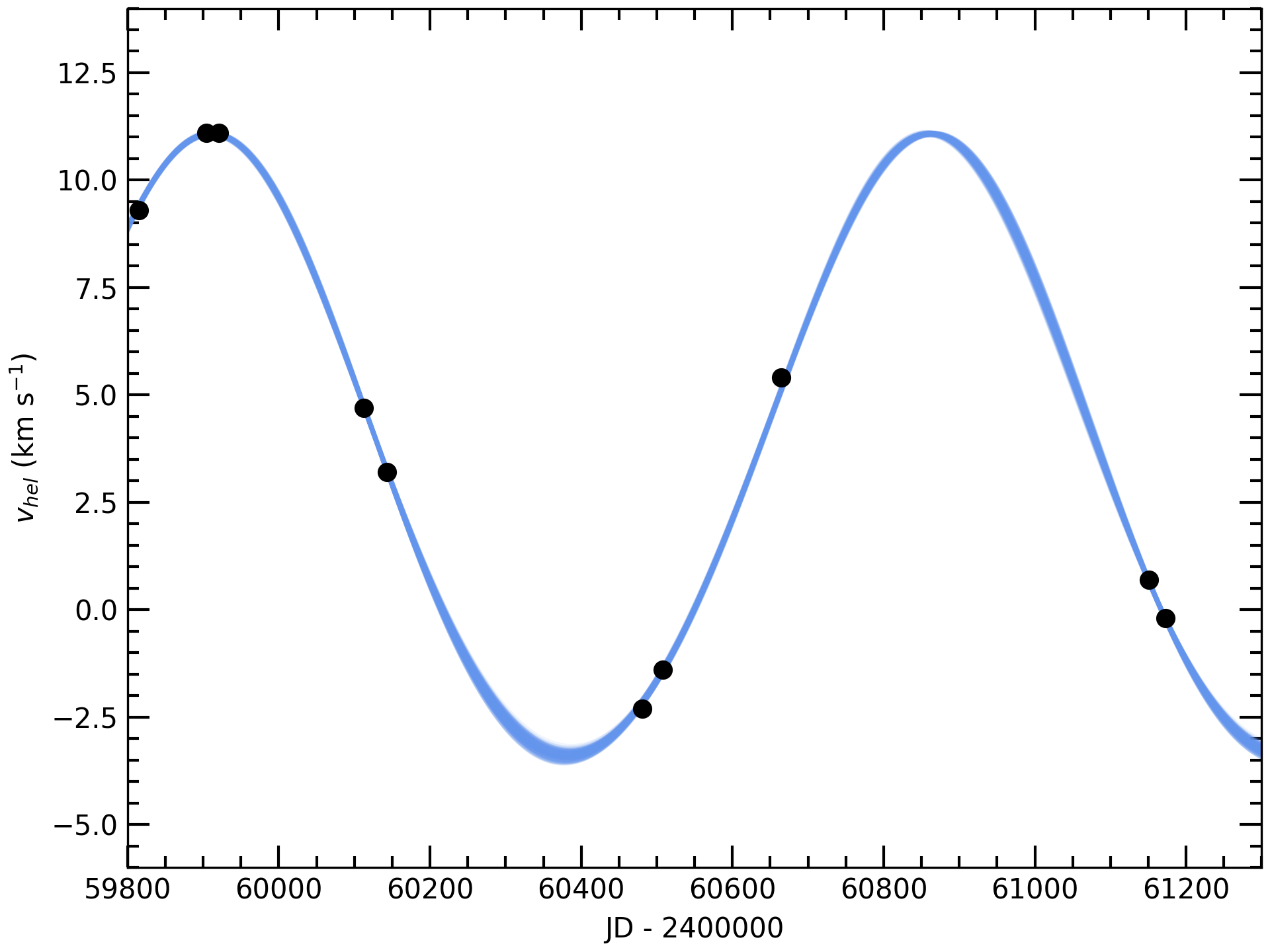}
    \caption{MCMC fit to the radial velocity curve of
      Gaia~DR3~6593763230249162112.  The black points are the RV
      measurements from \citet{elbadry24b} and MIKE and the blue
      curves are the 100 best-fitting solutions from the MCMC.  The
      orbit has a period of 952~d, a semi-amplitude of 7.3~\kms, and
      an eccentricity of 0.06.}\label{fig:6593rvs}
\end{figure}

\end{itemize}

\subsection{Candidates with Spectroscopic Binary Orbits}

The DR3 spectroscopic binaries in our sample largely contain
hotter, more massive primary stars than the astrometric binaries
described above.

\begin{itemize}

\item{Gaia~DR3~2000733415898027264 was classified in DR3 as a
  single-lined spectroscopic binary with a 5.0~\msun\ companion in a
  relatively short-period ($53.80 \pm 0.25$~d) orbit.  We obtained three APF
  spectra, which showed that the star is hot (A-type) and rapidly
  rotating.  The only feature from which we could measure a velocity
  was H$\beta$.  Over a 20~d span, we detected no change in the
  velocity of the star, with a precision of $\sim4$~\kms.  The
  measured velocity was also in agreement with the mean value from DR3
  within the uncertainties.  We rule out the DR3 orbital solution and
  conclude that the broad lines may have led to erroneous Gaia
  velocity measurements. }

\item{Gaia~DR3~2086448353089047808 was detected as an eclipsing binary
  by TESS \citep{tess} and therefore is not a candidate compact object
  binary.  We did not conduct spectroscopic follow-up.}

\item{Gaia~DR3~2226444358294583680 was classified in DR3 as an SB1
  consisting of a $\sim2$~\msun\ primary and a secondary with a
  lower mass limit of 29~\msun\ with a $97\pm1$~d period.  We obtained three APF
  spectra, which were consistent with an early A star.  The only
  feature from which we could measure a velocity was H$\beta$.  We
  found modestly significant evidence ($p = 0.08$) for RV variations
  at the $\sim10$~\kms\ level over the 13~d covered by our
  observations.  However, the high mass of the putative companion star
  predicts a velocity amplitude of $K_{1} = 164$~\kms.  Our velocity
  measurements are strongly inconsistent with this orbit.  The APF
  velocities are also $>30$~\kms\ away from the mean DR3 velocity,
  although the unusually large uncertainty of the Gaia velocity means
  that this offset is not statistically significant.  This system
  could be a binary, but it does not appear to contain a compact
  object.}

\item{Our first spectrum of Gaia~DR3~4060365702574410752 revealed that
  it is an SB2 with the secondary exhibiting only slightly weaker
  lines than the primary.  The velocity separation between the two
  stars was $\sim90$~\kms, accounting for most of the
  rv\_amplitude\_robust value of 116~\kms\ in the DR3 catalog.  Since
  this system clearly contains two luminous stars, we did not follow
  it up any further.}

\item{In the DR3 spectroscopic binary catalog,
  Gaia~DR3~5625482713303185408 was listed as a binary with a period of
  $687\pm16$~d, a primary mass of $1.74^{+0.05}_{-0.06}$~\msun, and a secondary of $>6.8$~\msun.
  We obtained a total of ten spectra of this star, seven with MIKE and
  three with IMACS, spanning 732~d.  We classified the primary star as
  A7--F0, with a rotation velocity of 270~\kms.  Although our velocity
  measurements only have a typical precision of 5~\kms\ because of the
  high temperature and linewidth, we did not find any evidence for
  radial velocity variability.  A $\chi^{2}$ test of the hypothesis
  that the velocity is constant results in a $p$-value of $p=0.85$,
  indicating that we cannot reject the hypothesis.  Since these
  observations span the full Gaia period, we rule out a binary system
  with properties similar to those determined in DR3.  We suggest that
  the broad lines may have led to erroneous Gaia velocity measurements
  and conclude that Gaia~DR3~5625482713303185408 does not have a
  massive companion.}

\item{Gaia~DR3~5846362195472084992 was classified in DR3 as an SB1
  binary with a period of $764\pm100$~d and a secondary mass above 10.5~\msun.  We obtained 14 MIKE spectra, from which we determine a spectral type of F0V via comparison to the similar star HD~10148 and the classification descriptions of \citet{gc09}.  The implied primary mass is therefore $\sim1.6$~\msun\ \citep{pecaut12}, in reasonable agreement with the DR3 mass of $1.74 \pm 0.06$~\msun.  We confirmed that the star is indeed a single-lined binary, but derived a period of $361.5^{+12.9}_{-9.7}$~d and a semi-amplitude of $K_{1} = 12.8^{+21.2}_{-5.6}$~\kms
  (see Fig.~\ref{fig:5846rvs}).  Because the orbital period is very close to 1~yr, the RV data provide phase coverage for only half of the orbit despite observations spanning four orbital cycles.  We suggest that aliasing of the nearly 1~yr period may have biased the Gaia orbital solution.   The lack of full phase coverage leaves the semi-amplitude and eccentricity quite uncertain.  The mass function is $f = 0.045^{+0.183}_{-0.032}$~\msun, so the range of possible companion masses remains rather wide ($M\sin{i} = 0.61^{+0.63}_{-0.23}$~\msun).  The companion is most likely either a low-mass main sequence star or a white dwarf, but we cannot currently rule out a higher mass secondary.  If the Gaia spectroscopic orbit were correct, the RUWE for this source should be $\sim15$, compared to the DR3 value of 1.09.  However, the RUWE is somewhat smaller than implied by our orbital solution as well, indicating that the inclination angle may be high.  This source is not likely to have received a successful astrometric orbital solution in DR3, in significant part because of the poor orbital coverage from the 1~yr period.}

\begin{figure}
    \centering
    \includegraphics[width=0.47\textwidth]{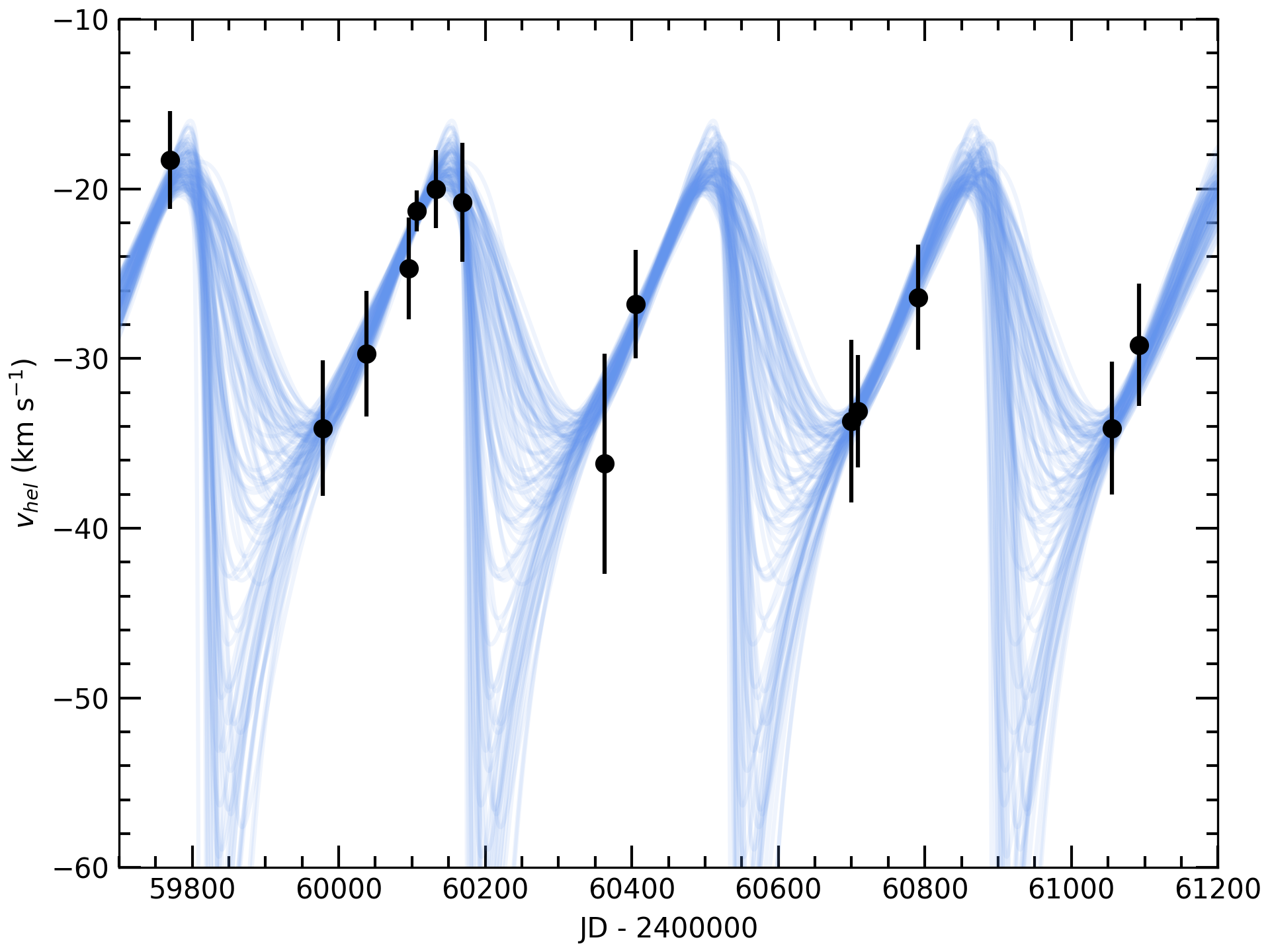}
    \caption{MCMC fit to the radial velocity curve of
      Gaia~DR3~5846362195472084992.  The black points are the RV
      measurements from MIKE and the blue curves are the 100
      best-fitting solutions from the MCMC.  The orbit has a period of
      362~d, a semi-amplitude of 12.8~\kms, and an eccentricity of
      0.51, although the amplitude and eccentricity are fairly
      uncertain because of the lack of coverage of the RV
      minimum.}\label{fig:5846rvs}
\end{figure}

\item{DR3 listed Gaia~DR3~6059985721200365184 as a binary with a $910\pm102$~d
  period and a $>11.8$~\msun\ dark secondary.  The goodness of fit
  value for the SB1 solution was quite small, but the significance was
  also low.  We obtained three MIKE spectra, from which it was clear
  that the star was quite hot (likely an early A spectral type) and
  rapidly rotating, consistent with the temperature of 9518~K
  estimated from Gaia spectrophotometry and the line broadening of
  262~\kms\ from the RVS spectra.  Over the 1367~d spanned by the MIKE
  spectra, we found no change in the radial velocity of the star with a
  precision of $\sim5$~\kms.  We therefore rule out the spectroscopic
  orbital solution.  The high RUWE value does indicate likely
  binarity, but the radial velocity amplitude of any binary orbit must
  be relatively small.}

\item{Gaia~DR3~6102598776102841344 was detected as an eclipsing binary
  by TESS and therefore is not a candidate compact object binary.  We
  did not conduct spectroscopic follow-up.}

\item{Gaia~DR3~5352109964757046528 had a $9.868\pm0.003$~d period in the DR3
  catalog and was classified as an SB1 system.  Our MIKE spectrum
  revealed at least four distinct components of the He absorption
  lines, confirming that the system contains multiple luminous stars.
  We did not continue to follow it to derive independent parameters.}

\end{itemize}

\subsection{Candidates with Combined Astrometric and Spectroscopic Binary Orbits}

\begin{itemize}

\item{Gaia~DR3~1864406790238257536 was classified as an A star with
  both astrometric and spectroscopic orbital solutions in DR3.  The
  orbital period is $876\pm59$~d and the inferred companion mass is
  $20^{+13}_{-8}$~\msun.  We obtained a Keck~HIRES spectrum in which it was clear
  that the star is actually an SB2 binary containing one component
  with broad lines and another with narrow lines.  We therefore did
  not follow it further to determine an independent spectroscopic
  orbit.  It seems reasonable to assume that the DR3 orbital period
  is correct given the agreement between the astrometric and
  spectroscopic parameters, but the detection of a second luminous
  component invalidates the masses determined for the AstroSpectroSB1 solution.
  This system does not contain a black hole.}

\item{Gaia~DR3~3263804373319076480 was classified by \citet{shahaf23}
  as an AMRF Class~III system with a secondary potentially in the mass
  gap regime at $M_{2} = 2.75$~\msun\ (however, the DR3 catalog lists a secondary mass of $M_{2} = 0.99^{+0.34}_{-0.26}$~\msun).  The Gaia period is $510.7\pm4.7$~d, with a moderate eccentricity.  Based on
  follow-up spectroscopy with FEROS and MIKE, combined with archival velocity measurements from LAMOST and APOGEE, we found that the period is somewhat longer than determined by Gaia, at $605.9^{+1.1}_{-1.3}$~d (see Fig.~\ref{fig:3263rvs}).  The semiamplitude is $14.4^{+0.7}_{-0.5}$~\kms\ and the eccentricity is $0.26\pm0.02$.  Although the difference from the Gaia period is only 3 months, the period measurements disagree at high significance.  Since the Gaia orbit is an AstroSpectroSB1 solution, with consistent spectroscopic and astrometric fits, and the star is relatively cool (5613~K), it is surprising that the DR3 result is wrong.  We note that the rv\_amplitude\_robust value for the star is 29.6~\kms, compared to the 63~\kms\ amplitude for the AstroSpectroSB1 solution.  DR3 spans approximately two orbital periods of the system, so unless the sampling of the Gaia velocities was quite unlucky, such a small value for the observed amplitude is unexpected.  In contrast, the amplitude of our orbital solution matches rv\_amplitude\_robust within the uncertainties.  The RUWE value is also consistent with our orbit and inconsistent with the Gaia orbit.  We compute a mass function of $f = 0.17\pm0.02$~\msun, indicating a minimum companion mass of 0.72~\msun.  The secondary mass is very close to the primary mass of 0.77~\msun, so if the secondary were luminous its lines should have been easily visible in our spectra.  Given the lack of such a signal, we conclude that the companion is a compact object, most likely a white dwarf.  However, if the inclination is less than $\sim30\degr$, the companion mass would be above the Chandrasekhar mass and a neutron star would be required.  We calculated the expected RUWE as a function of inclination angle for the correct orbital solution using gaiamock \citep{elbadry_forwardmodel} and found that the inclination angle must be above $\sim50\degr$ to match the observed RUWE, suggesting a true secondary mass of no more than $\sim0.9$~\msun.}

\begin{figure}
    \centering
    \includegraphics[width=0.47\textwidth]{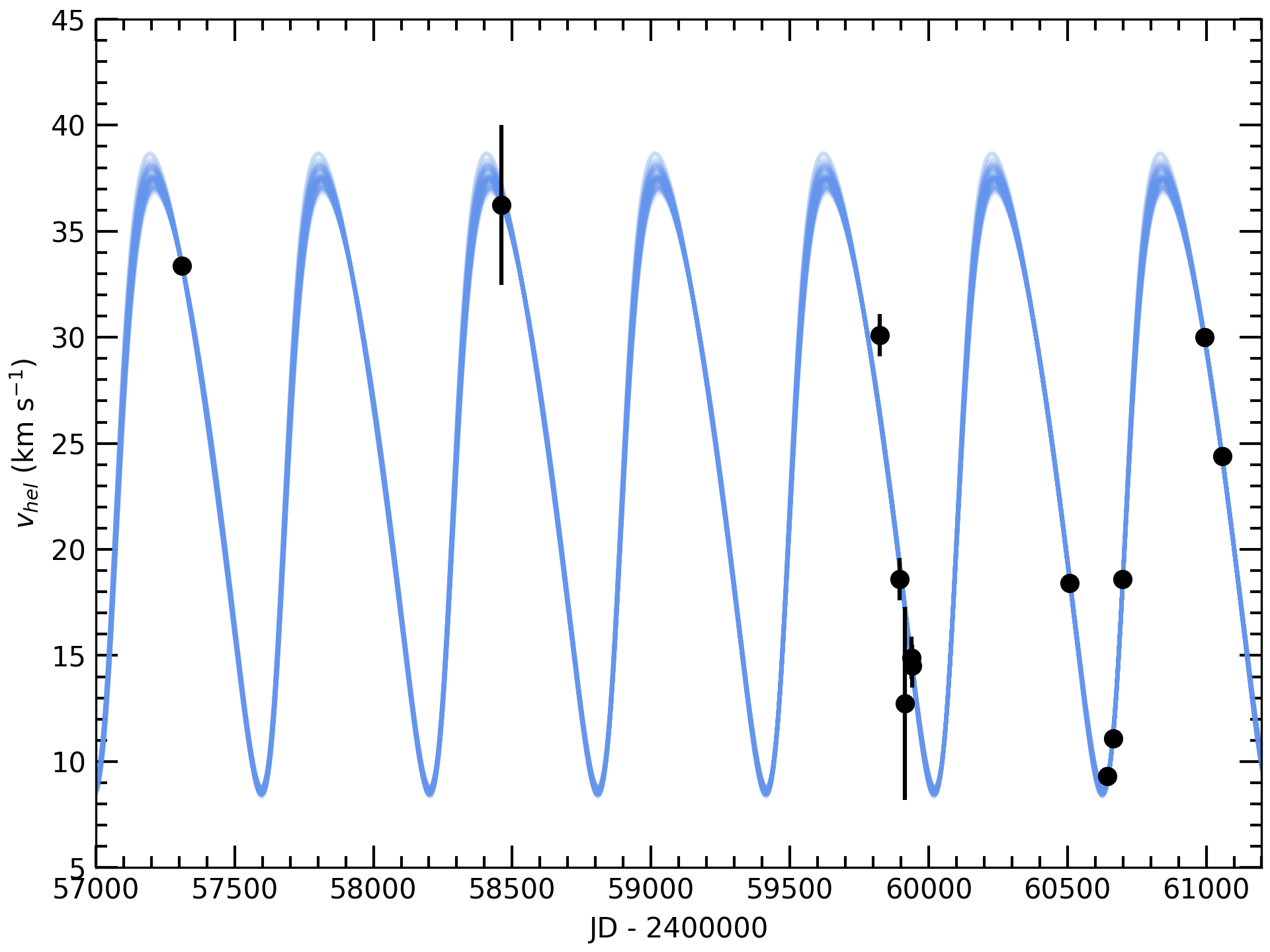}
    \caption{MCMC fit to the radial velocity curve of
      Gaia~DR3~3263804373319076480.  The black points are the RV
      measurements from MIKE, FEROS, and the literature and the blue curves are the 100
      best-fitting solutions from the MCMC.  The orbit has a period of
      606~d, a semi-amplitude of 14.4~\kms, and an eccentricity of
      0.26.}\label{fig:3263rvs}
\end{figure}

\end{itemize}

\tabletypesize{\scriptsize}
\begin{deluxetable*}{l c c c c c c c}
\tablecaption{RV Orbital Solutions\label{tab:new_orbits}}
\tablehead{source\_id & $v_{0}$ & $P$ & $K_{1}$ & $e$ & $\omega$ & T$_{peri}$ & $f$ \\ 
  & (\kms) & (d) & (\kms) & & & & (\msun)}
\startdata
\sidehead{\textit{Gaia astrometric binaries}}
   5593444799901901696 & $12.54^{+0.62}_{-0.66}$ & $20.023^{+0.022}_{-0.025}$  & $9.4^{+0.6}_{-0.5}$  & $0.14^{+0.09}_{-0.07}$  & $1.57^{+0.58}_{-0.99}$  & $2459919.6^{+1.6}_{-2.8}$ & $0.0016^{+0.0004}_{-0.0003}$ \\
   3509370326763016704 &  $-19.14^{+0.08}_{-0.17}$ & $457.88^{+9.63}_{-6.46}$ & $1.9^{+2.0}_{-0.3}$ & $0.54^{+0.21}_{-0.11}$ & $-2.64^{+0.18}_{-0.20}$ & $2459553.2^{+8.4}_{-12.2}$ & $0.00021^{+0.00057}_{-0.00005}$ \\
   3640889032890567040 &  $22.98^{+0.20}_{-0.19}$ & $781.0^{+8.7}_{-7.7}$ & $8.2 \pm 0.3$ & $0.34 \pm 0.02$ & $2.43^{+0.04}_{-0.05}$ & $2459413.9^{+12.0}_{-13.0}$ & $0.037^{+0.004}_{-0.003}$ \\
   6281177228434199296 & $-3.87 \pm 0.08$  & $273.40^{+0.19}_{-0.18}$  & $15.94^{+0.12}_{-0.11}$  & $0.553 \pm 0.008$ & $2.97\pm0.01$  & $2459692.7^{+0.5}_{-0.6}$ & $0.066 \pm 0.001$ \\
   6802561484797464832 & $-51.07^{+0.13}_{-0.27}$  & $524.8^{+5.4}_{-5.6}$  & $2.6^{+0.8}_{-0.2}$  & $0.46^{+0.07}_{-0.04}$  & $2.81\pm0.12$  & $2460067.6^{+15.1}_{-22.4}$ & $0.00073^{+0.00067}_{-0.00017}$ \\
   6588211521163024640 & $-13.28 \pm 0.07$  &  $641.4^{+4.6}_{-4.1}$  & $4.84^{+0.05}_{-0.04}$  &  $0.30 \pm 0.01$  & $1.81 \pm 0.06$  &  $2460004.7^{+7.8}_{-7.9}$ & $0.0065 \pm 0.0002$ \\
   6601396177408279040 & $-22.81 \pm 0.06$  & $525.5^{+8.9}_{-7.9}$  & $1.46^{+0.10}_{-0.09}$  & $0.26^{+0.05}_{-0.06}$  & $-2.95\pm0.24$  &  $2459693.7^{+24.1}_{-24.9}$ & $0.00015^{+0.00003}_{-0.00002}$ \\
   6593763230249162112 & $3.46 \pm 0.06$  & $951.5^{+6.3}_{-7.5}$  & $7.3 \pm 0.2$  & $0.06^{+0.02}_{-0.01}$ & $0.19^{+0.60}_{-0.47}$  & $2459937^{+87}_{-67}$ & $0.037^{+0.003}_{-0.002}$ \\
\hline
\sidehead{\textit{Gaia SB1 binaries}}
   5846362195472084992 &  $-26.8^{+3.6}_{-3.9}$ & $361.5^{+12.9}_{-9.7}$ & $12.8^{+21.2}_{-5.6}$ & $0.51^{+0.32}_{-0.37}$  & $1.75^{+0.71}_{-0.92}$  & $2459878^{+62}_{-64}$ & $0.045^{+0.183}_{-0.032}$ \\
\hline
\sidehead{\textit{Gaia AstroSpectroSB1 binaries}}
   3263804373319076480 &  $24.86^{+0.63}_{-0.56}$  & $605.9^{+1.1}_{-1.3}$ & $14.4^{+0.7}_{-0.5}$ & $0.26\pm0.02$  & $-2.10^{+0.18}_{-0.19}$ & $2457053.0^{+8.4}_{-8.7}$ & $0.17 \pm 0.02$ \\
\hline
\sidehead{\textit{Gaia accelerating sources}}
   3689209059942075008 &  $-17.06 \pm 0.07$ & $791.93 \pm 1.16$ & $19.34 \pm 0.09$ & $0.537\pm0.004$ & $2.82\pm0.01$ & $2456584.8^{+5.6}_{-5.5}$ & $0.356 \pm 0.006$ \\
   5937774698871600512 &  $-2.9^{+0.7}_{-0.8}$ & $1122^{+291}_{-105}$ & $5.1^{+3.8}_{-1.2}$ & $0.35^{+0.47}_{-0.26}$ & $1.41^{+0.76}_{-1.41}$ & $2460559^{+123}_{-254}$ & $0.011^{+0.012}_{-0.005}$ \\
\enddata
\end{deluxetable*}

\subsection{Accelerating Sources}

\begin{itemize}

\item{Gaia~DR3~3689209059942075008 had both highly significant
  accelerations and acceleration derivatives with time in the DR3
  catalog, suggesting a period within a factor of a few of the DR3
  duration.  Through extensive spectroscopic followup with both APF
  and Magellan, we determined an orbit with a period of $791.93\pm1.16$~d, a
  semi-amplitude of $K_{1} = 19.34 \pm 0.09$~\kms, and an eccentricity of $0.537 \pm 0.004$ (Fig.~\ref{fig:3689rvs}).
  The binary mass function is $f=0.356 \pm 0.006$~\msun, giving a minimum mass
  for the companion of M~$\sin{i} = 1.16$~\msun.  This mass is quite close to the
  range of 1.25--1.4~\msun\ for most of the neutron stars in
  astrometric binaries identified by \citet{elbadry24b}.  The companion may therefore
  be either an ultramassive white dwarf or a neutron star.  A determination of the
  actual inclination in Gaia DR4 will conclusively determine the
  nature of the companion.  For now, we note that the RUWE value is modestly lower than would be expected for the derived orbit, suggesting that the inclination is likely close to edge-on ($i \gtrsim 75\degr$).  The true secondary mass would then be 1.2~\msun.  According to both gaiamock and the \citet{lam25} method, this binary should have had a moderately high probability ($>60$\%) of being included in the DR3 orbital catalog.  No binary systems with these properties should have received only an acceleration solution.  Without access to the epoch astrometry measurements, we cannot assess why processing of this source stopped in the variable acceleration portion of the processing cascade \citep[see][]{babusiaux23, halbwachs23}.}

\begin{figure}
    \centering
    \includegraphics[width=0.47\textwidth]{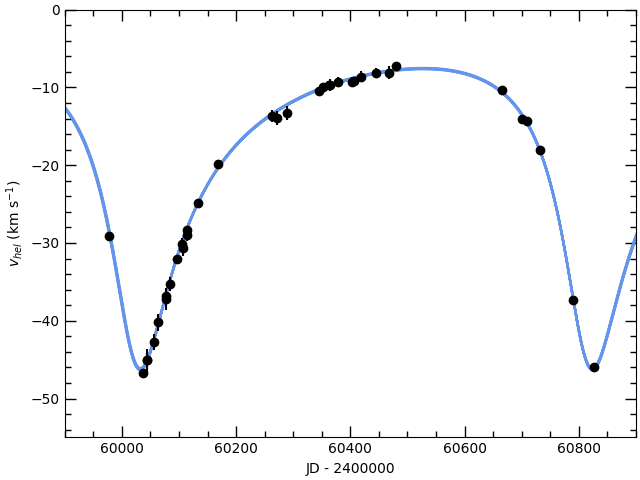}
    \caption{MCMC fit to the radial velocity curve of
      Gaia~DR3~3689209059942075008.  The black points are the RV
      measurements from MIKE and APF and the blue curves are the 100
      best-fitting solutions from the MCMC.  The orbit has a period of
      792~d, a semi-amplitude of 19.3~\kms, and an eccentricity of
      0.54.  The minimum companion mass is 1.16~\msun, so the companion star is certainly a compact object, most likely a neutron star.}\label{fig:3689rvs}
\end{figure}

\item{Gaia~DR3~1873093722367193216 exhibited a varying
  acceleration with time in the DR3 catalog.  Gaia photometry suggests
  that the star is evolved and relatively hot.  We obtained four
  epochs of APF spectroscopy spanning 79~d.  Measuring accurate
  velocities was challenging because of the low S/N of the APF data and
  the star's broad lines.  We detected a modestly significant velocity
  change of 9~\kms\ over this time interval, which is consistent with
  the star being a genuine binary but does not significantly constrain
  the period or amplitude.  }

\item{Gaia~DR3~5358111034809618304 had the highest significance among
  the accelerating sources in our sample.  It received a 7-parameter
  astrometric acceleration solution and a second-order RV acceleration
  solution in DR3.  Our first spectrum showed two narrow blended
  components with similar line strengths for most spectral lines.  We
  classified this star as an SB2 binary and did not follow it further.}

\item{Gaia~DR3~409488405416253696 was fit with a 7-parameter
  astrometric acceleration solution and a second order RV acceleration
  in DR3.  The significance of the RV acceleration was lower than for
  some of the other stars in our sample.  The total RV amplitude
  reported in DR3 was 96~\kms.  We obtained six epochs of APF
  spectroscopy spanning 158~d.  Measuring accurate velocities was
  challenging because of the low S/N of the APF data and the star's
  broad lines.  We did not detect significant velocity changes and the
  measured velocities are consistent with the mean velocity reported
  in DR3.}

\item{Gaia~DR3~4169598884257076864 was best fit by constant
  accelerations both along the line of sight and in the plane of the
  sky, with a DR3 RV amplitude of 74~\kms.  The Gaia accelerations are
  highly significant.  From our spectroscopy, we classified the star as
  an early F star featuring relatively rapid rotation.  We obtained 8 MIKE spectra and one IMACS spectrum spanning a total of 1074~d.  Although the velocity was consistent with remaining constant for the first three months of observations, continued monitoring revealed a velocity range of at least 16~\kms\ on longer time scales, confirming that the star is indeed a binary.  Fits to the velocity data with TheJoker indicated that the period is likely longer than 1200~d, although a few solutions at $P\sim500$~d also exist.  The semi-amplitude of the binary is most likely $K_{1} \sim 10$\kms, but larger values are not ruled out, especially at longer periods.  These results suggest a mass function of $f\sim0.2$~\msun, but significantly larger values are also possible.  Further monitoring will be needed to determine the nature of the binary system.}

\item{Gaia~DR3~5939087648856671232 had a 9-parameter astrometric
  acceleration solution and a linear RV acceleration in the DR3
  catalog.  The significance of the RV acceleration was very high,
  whereas the significance of the astrometric acceleration terms was
  somewhat lower.  However, the latter may be affected by the large
  derivatives for the astrometric accelerations.  This star also had
  the largest DR3 RV amplitude of the accelerating sources in our
  catalog (209~\kms).  Spectroscopic follow-up with MIKE showed that
  the star is a late A star with a rotation velocity of
  $\sim200$~\kms.  Like Gaia~DR3~4169598884257076864, our first six
  months of RV monitoring showed no significant variability and a
  velocity close to that reported in DR3, but another observation
  eight months later detected a velocity change of $\sim18$~\kms.  After another 1.5~yr, the velocity returned to approximately its original value.  We
  found a lower limit on the period of $\sim500$~d, with the most likely solution around 900~d.  The binary semi-amplitude is probably $K_{1} \approx 10$~\kms, but larger amplitudes are not ruled out.  The mass function may therefore be in the range of $f\sim0.09$~\msun.}

\item{Gaia~DR3~392363030772060672 had a quadratic radial velocity fit
  and a seven-parameter astrometric acceleration fit.  The first
  derivative of the radial velocity and the acceleration in right
  ascension were only modestly significant, but the second derivative
  of the RV and the acceleration in declination were much better
  determined.  Gaia spectrophotometry indicated a somewhat reddened
  late A star, and the linewidth measured from the RVS spectra was
  146~\kms.  We obtained four epochs of APF spectroscopy spanning
  92~d.  Measuring accurate velocities was challenging because of the
  low S/N of the APF data and the star's broad lines.  We did not
  detect significant changes in the velocity over this time span, but
  cannot draw strong conclusions about binarity given the short time
  baseline and large velocity uncertainties.}

\item{Gaia~DR3~2044799672971280384 received a quadratic radial
  velocity fit and a nine-parameter astrometric acceleration solution.
  The astrometric accelerations were significant at $>10\sigma$, as was
  the acceleration derivative in declination.  The star is at
  relatively low Galactic latitude ($b=5.8\degr$), but 3-D dust maps
  indicate that the expected reddening at the astrometric distance is
  small, so that the observed color implies a solar-like main sequence
  star.  We obtained three APF spectra spanning 51 days, from which we
  classified the star as spectral type F8/9, with slightly broadened
  lines indicating a projected rotation velocity of 16~\kms.  We did
  not detect a statistically significant change in velocity over the
  course of these observations, although the expected acceleration in
  51 days from the DR3 catalog is only 1.5~\kms, which is consistent with our measurements.  The mean velocity from the APF spectra was 6~\kms\ away from
  the value of $22.0 \pm 3.7$~\kms\ reported in DR3, but given the
  uncertainty this difference is of modest significance.  Since
  Gaia~DR3~2044799672971280384 is a relatively cool star, there is no
  obvious reason why its RVs would not have been measured accurately
  from the RVS data, and hence we assume that its reported velocity
  amplitude of 52~\kms\ is real.  However, we are unable to provide
  additional positive evidence in favor of binarity without RV
  follow-up over a longer time baseline.}

\item{Gaia~DR3~5309744205505321600 was fit with a seven-parameter
  astrometric acceleration solution and a linear acceleration of its
  radial velocity in DR3.  The RV acceleration term was quite large,
  0.15~\kms~d, indicating that observations over even a few months
  should detect a velocity change.  We obtained eight spectra (six
  with MIKE, two with IMACS) covering 429~d.  From the MIKE data, we
  estimated a spectral type of A3 and a rotation velocity of
  $\sim200$~\kms.  Although the high temperature and broad lines made
  velocities difficult to determine, we did not detect any significant
  RV variability.  From a $\chi^{2}$ test, we found a $p$-value of
  $p=0.33$ for the constant-velocity hypothesis.  The lack of
  spectroscopic evidence for binarity does not contradict the Gaia
  astrometric acceleration, but suggests that a massive companion is
  unlikely.}

\item{Gaia~DR3~5937774698871600512 was fit with variable acceleration
  models in both radial velocity and astrometry.  We obtained 14
  epochs of MIKE spectroscopy across 1173~d, from which we determined a
  spectral type of F0 and a rotation velocity of $v\sin{i} =
  93$~\kms.  The radial velocity curve covers a full orbital cycle, with a best-fit period of $1122^{+291}_{-105}$~d.  The
  velocity semi-amplitude is constrained to be $5.1^{+3.8}_{-1.2}$~\kms\ (see Fig.~\ref{fig:5937rvs}).  The uncertainties on the orbital parameters are relatively large because of the small amplitude and the increased velocity uncertainties resulting from the high rotation velocity.  However, the data are sufficient to rule out a massive companion unless the binary is seen close to face-on.  The minimum companion mass is 0.35~\msun. }

\begin{figure}
    \centering
    \includegraphics[width=0.47\textwidth]{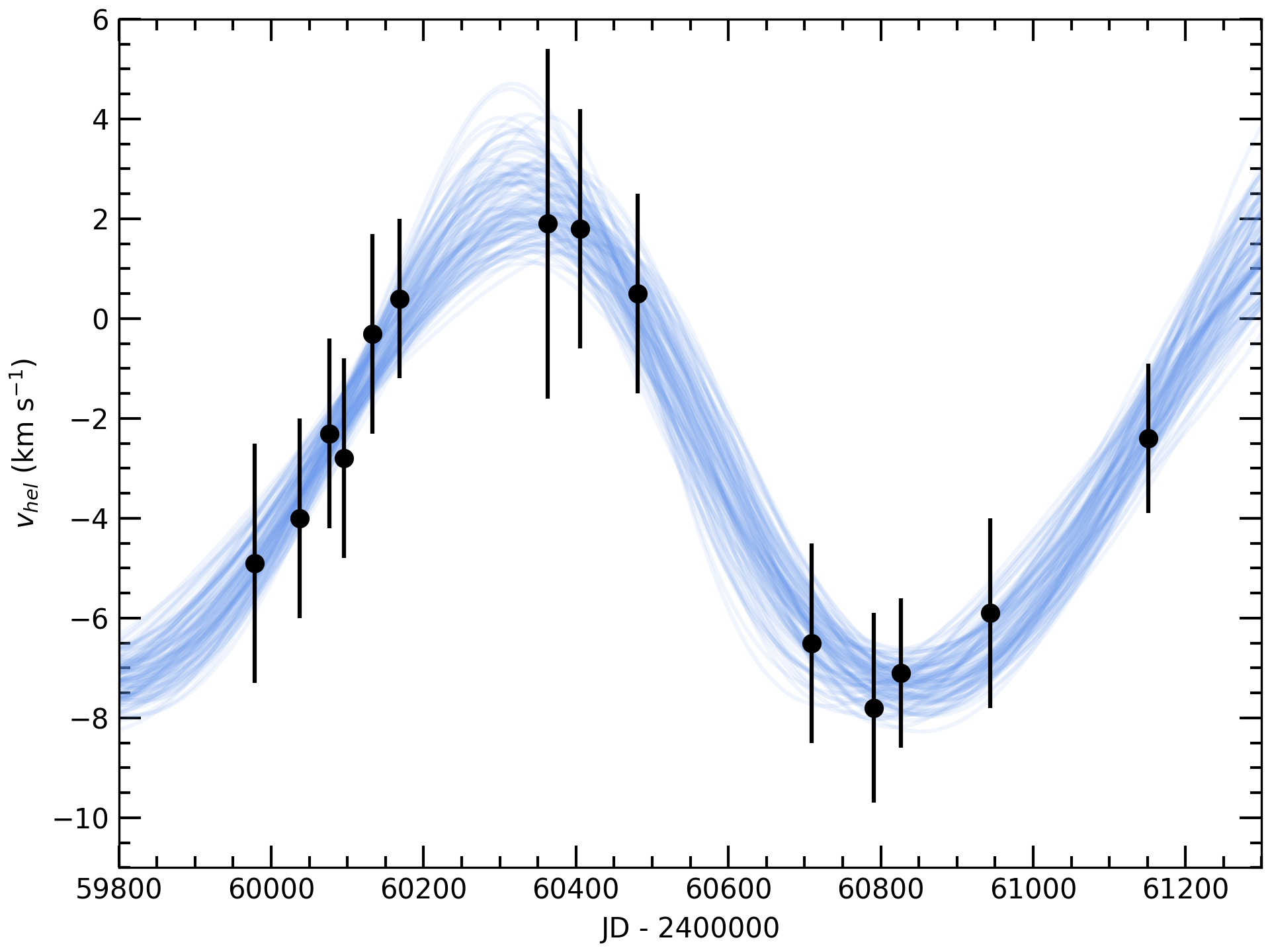}
    \caption{MCMC fit to the radial velocity curve of
      Gaia~DR3~5937774698871600512.  The black points are the RV
      measurements from MIKE and the blue curves are the 100
      best-fitting solutions from the MCMC.  The orbit has a period of
      1122~d, a semi-amplitude of 5.1~\kms, and an eccentricity of
      0.35.}\label{fig:5937rvs}
\end{figure}

\item{Gaia~DR3~5530445257521690624 was fit with a constant radial
  velocity acceleration with modest significance ($\sim5\sigma$) along
  with a variable astrometric acceleration.  Its DR3 velocity
  amplitude of 50.14~\kms\ just barely exceeded our selection
  threshold.  We obtained 18 MIKE spectra, 3 FEROS spectra, and 1 IMACS spectrum, covering a total of 1104~d, from which we
  measured its stellar parameters and velocities.  We classified the
  star as a mildly metal-poor F9 star.  Over the course of our
  observations, we measured a nearly linear velocity change of
  $\sim13$~\kms, which reached a minimum some time in the second half of 2025 and then began increasing.  We constrained the orbital period to be longer than
  $\sim1300$~d, consistent with expectations for accelerating sources
  in DR3 \citep{elbadry_forwardmodel,lam25}.  The minimum semi-amplitude is $\sim6$~\kms, but the data allow significantly larger amplitudes, especially at longer periods.  If the semi-amplitude is larger than $\sim12$~\kms\ (which occurs at periods longer than 3000~d), then the minimum secondary mass would exceed the Chandrasekhar mass.  Our period constraint therefore requires a neutron star or black hole companion if the Gaia rv\_amplitude\_robust value is correct.  However, the DR3 radial velocity of $v_{\rm hel} = 22.8 \pm 2.6$~\kms\ and the reported rv\_amplitude\_robust together imply a velocity range of $-2$~\kms\ to 48~\kms\ (unless the eccentricity is quite high), which is inconsistent with our measured minimum velocity of $v_{\rm hel} = 14$~\kms\ (see Fig.~\ref{fig:5530rvs}).  The true velocity amplitude is therefore unclear, but Gaia~DR4 should provide additional insight.  We also found that the star was significantly enriched in $s$-process elements, with $\mbox{[X/Fe]} > 1$ for Ba, Ce, Nd, and Y.  This chemical abundance pattern likely originated via mass transfer from a companion star that evolved through the asymptotic giant branch phase, suggesting that the secondary is probably a white dwarf \citep[e.g.,][]{yamaguchi26}.}

\begin{figure}
    \centering
    \includegraphics[width=0.47\textwidth]{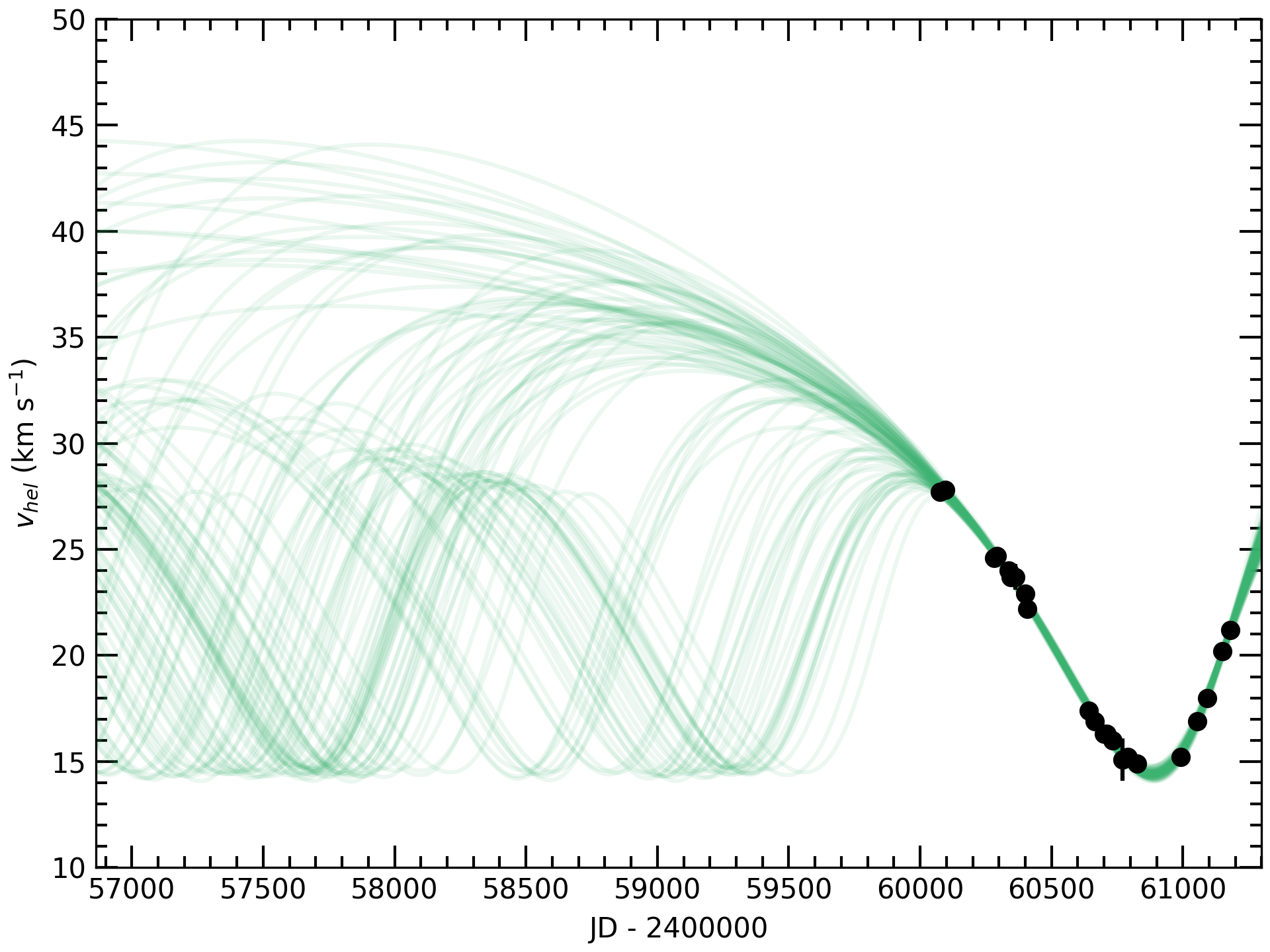}
    \caption{TheJoker fits to the radial velocity curve of
      Gaia~DR3~5530445257521690624.  The black points are the RV
      measurements from MIKE, FEROS, and IMACS, and the green curves are 100
      possible orbital solutions.  The orbit has a period of 
      $\gtrsim1300$~d and a semi-amplitude of $\gtrsim6$~\kms.}\label{fig:5530rvs}
\end{figure}

\end{itemize}

\section{Discussion}
\label{sec:discussion}

\subsection{Summary of Spectroscopic Follow-up Results}

Out of 23 objects with astrometric and/or spectroscopic orbits
indicating possible compact object companions, we confirmed that 18
are indeed binaries or multiple systems.  Four of the exceptions are hot,
very rapidly rotating stars for which the Gaia radial velocities were
apparently erroneous (but one of these does show tentative evidence for low-amplitude RV 
variations).  The other star for which we did not detect RV variation or a spectroscopic secondary is the faintest star in our sample, and is among the faintest $\sim0.1$\%\ of the DR3 astrometric binary catalog overall, perhaps suggesting spurious astrometry or underestimated uncertainties.

Although three objects originally selected for
our sample (Gaia~BH1, Gaia~BH2, and Gaia~NS1) were previously verified
as containing black holes or neutron stars, the DR3 orbital solutions
for all other binaries with reported dark companions above
2~\msun\ are wrong.  Adopting a minimum significance of 10 and a
maximum goodness\_of\_fit of 4 would retain the known genuine compact
objects while excluding all but one of the false solutions.  This
goodness\_of\_fit threshold is much more conservative than the $F_{2}
< 25$ criterion suggested by \citet{halbwachs23}.  The proposed
significance limit is not very different from the minimum of
$\texttt{significance} = 12$ that \citeauthor{halbwachs23} used for
automatic acceptance of orbital solutions, but is higher than the
value of 5 set as the minimum value for inclusion in the DR3 catalog.
For objects selected from the extreme tails of the binary catalog (or
perhaps more generally if one prefers to prioritize sample purity over
completeness), these more stringent cuts may be helpful in identifying
reliable orbital solutions.  Hot stars with apparently large radial velocity variations in the DR3 data should be treated with caution.

We also followed up 11 accelerating sources, of which 7 were confirmed as
binaries.  Three of the remaining four stars were hot, and three were
observed only with APF (which is mounted on a small telescope) because
of their northern declinations.  The lack of RV variability for these
stars likely results from a combination of poor Gaia RV measurements
for stars with broad lines and limited time coverage of our
spectroscopic campaign for some targets.  The stellar temperature and
rotation velocity should have little to no effect on the Gaia
astrometry, so the astrometric accelerations are presumably reliable.
However, massive companions should have been detected by our velocity
measurements unless the periods are very long.  We therefore conclude
that the stars without detected RV changes are unlikely to host
compact objects.  We look forward to Gaia DR4 astrometry to shed more
light on these candidate binary systems.

Among the accelerating targets, we found that the periods are long
($\gtrsim2$~yr), as expected for sources lacking DR3 orbital
solutions.  The only star for which we have been able to determine a
complete spectroscopic orbit so far does have a compact object companion 
with a mass of at least 1.16~\msun, implying an ultramassive white dwarf 
or neutron star.  Several other systems could plausibly contain companions
with $M_{2} \gtrsim 1$~\msun, and just one is likely inconsistent with
a massive companion.\footnote{Here we are assuming that none of the
  binaries are viewed close to face-on, such that the dependence of
  the velocity amplitude on $M_{2}\sin{i}$ makes a massive secondary
  appear significantly less massive.}  Stars with large accelerations
but no orbital solutions in DR3 therefore provide a promising pool of
candidates for additional wide binaries with compact objects, but
patience is needed to investigate these systems.

\subsection{Implications for the Population of Black Holes Detectable in DR3 and Beyond}

Prior to DR3, predictions for the number of binary systems containing a black hole and a luminous companion star that could be detected by Gaia varied by orders of magnitude \citep[e.g.,][]{breivik17,ml17,yamaguchi18,wiktorowicz20,chawla22,janssens22}.  Gaia~BH1 and BH2 were identified and confirmed relatively quickly \citep{elbadry23a,chakrabarti23,tanikawa23,elbadry23b}, while the other candidates that initially appeared the most promising were ruled out \citep{ebr22,elbadry23a}.  Here, we firmly exclude the possibility that any other published candidates from DR3 contain a black hole, indicating that the number of black hole-luminous companion binaries with DR3 orbital solutions is two.  With a revised understanding of the selections that went into the DR3 binary catalogs, \citet{chawla25} now conclude that the expected number to be detectable in DR3 is zero, which is approximately consistent with the observed result given Poisson uncertainties.

Nevertheless, the expected number of black hole binaries that Gaia will reveal with the longer time baseline and better orbital modeling in DR4 and DR5 remains in the range of dozens to hundreds \citep{shikauchi22,chawla25, nagarajan25b}.  The long orbital period and very low metallicity of Gaia~BH3 point to one possible path for identifying additional such systems, implying that the formation of stellar-mass black holes may be enhanced at low metallicities \citep{elbadry24c}.  \citet{nagarajan25} and \citet{lam25b} have explored the approach of specifically targeting metal-poor stars as potential black hole companions, identifying several promising DR3 candidates, but none have been confirmed yet.  \citet{mh25} used a selection based on large RUWE values (implying astrometric motion) and large rv\_amplitude\_robust values (implying RV variability) to uncover a new sample of candidate BH binaries without DR3 orbital solutions, but few of them have been followed up thus far.  Simply obtaining RV confirmation of the highest mass function solutions in the DR4 catalog, and/or refitting the DR4 epoch astrometry and velocities independently, are also likely to yield results, although if the periods of the remaining black hole binaries are longer than BH1 and BH2 then confirmation may take longer as well.  For candidates bright enough to have epoch RVs in DR4, a small number of independent follow-up velocities that agree with both the astrometric orbit and the Gaia RV curve may be sufficient to quickly demonstrate that the companion is a black hole.  On the other hand, fainter candidates without existing velocity data will likely require new RV measurements covering a substantial fraction of the orbit to provide confidence that the astrometric orbital solution is correct \citep[e.g.,][]{elbadry23b}.

\section{Conclusions}
\label{sec:conclusion}

We have presented an investigation into the nature of binary systems in the Gaia DR3 catalog with candidate non-luminous companions massive enough to be neutron stars or black holes.  We obtained spectroscopy of a complete sample of Gaia binaries classified as containing a dark secondary above 2~\msun.  Apart from the previously identified objects Gaia~BH1, Gaia~BH2, and Gaia~NS1, we found no new neutron stars or black holes.  The Gaia orbits for the other 20 stars in the sample are incorrect.  We confirmed that 75\%\ of these stars are indeed binaries, but their RV orbital solutions have periods that are not consistent with Gaia (with one exception), and their velocity semi-amplitudes are much smaller than predicted.  Several of these systems contain luminous secondaries, one hosts a likely white dwarf, and the rest have brown dwarf or M dwarf companions.  The remaining 25\%\ of the sample consists of stars that are hot, where the Gaia velocities are apparently inaccurate, or faint, where the Gaia astrometry may have larger uncertainties.  Stricter quality cuts on the Gaia orbital solutions ($\texttt{significance} > 10$, $F_{2} < 4$) can remove most of the erroneous solutions while retaining the confirmed compact objects.  For astrometric orbital solutions, these cuts would result in a largely clean sample, but if SB1 solutions are included as well then the sample purity would be 33\%.

We also obtained follow-up spectroscopy of a sample of 11 stars
selected from the Gaia DR3 catalog of accelerating sources to have
measured accelerations both along the line of sight and in the plane
of the sky, as well as large RV amplitudes.  We confirmed 63\%\ of
these stars as binaries; as in the case of the stars with orbital
solutions, the ones that do not exhibit RV changes are largely hot
stars.  The binaries appear to have periods longer than $\sim2$~yr,
but we have only been able to determine a complete orbit for two
sources thus far.  The secondary in one of those cases has a minimum mass of 1.16~\msun, indicating that it is either an ultramassive white dwarf or a neutron star.  

Based on these results, we conclude that only two wide binaries with black hole secondaries are contained in the Gaia DR3 binary catalogs.  There may be additional black holes in the DR3 acceleration catalogs, as previous follow-up of those sources has been limited.  However, since our sample includes the stars with the largest RV amplitudes, any black holes with smaller RV amplitudes must have periods longer than $\sim2$~yr for $M_{\mathrm{BH}} = 5$~\msun\ or longer than $\sim5$~yr for $M_{\mathrm{BH}} = 10$~\msun, similar to Gaia~BH3 \citep{panuzzo24}, necessitating long-term monitoring efforts.

Soon, the Gaia DR4 binary catalog will provide a significantly larger set of candidate wide binaries with black hole secondaries.  This study, along with the discovery of the first three Gaia black holes \citep{elbadry23a,elbadry23b, chakrabarti23,panuzzo24}, provides a guide to identifying these objects, separating them from contaminants, and confirming their nature.  In order to explore the population of black holes in wide binaries more completely, we encourage continued RV follow-up for DR3 accelerating sources and DR4 binaries when the next catalog is released.

\acknowledgments 
We thank Jhon Yana Galarza, Sasha Mintz, and Dan Kelson for assistance with MIKE observations, and Natsuko Yamaguchi for helpful conversations.  We thank the referee for suggestions that clarified our presentation of the results.

This work has made use of data from the European Space Agency (ESA) mission
{\it Gaia} (\url{https://www.cosmos.esa.int/gaia}), processed by the {\it Gaia}
Data Processing and Analysis Consortium (DPAC,
\url{https://www.cosmos.esa.int/web/gaia/dpac/consortium}). Funding for the DPAC
has been provided by national institutions, in particular the institutions
participating in the {\it Gaia} Multilateral Agreement.

This research has made use of NASA's Astrophysics
Data System Bibliographic Services.

Some of the data presented herein were obtained at Keck Observatory, which is a private 501(c)3 non-profit organization operated as a scientific partnership among the California Institute of Technology, the University of California, and the National Aeronautics and Space Administration. The Observatory was made possible by the generous financial support of the W. M. Keck Foundation.  The authors wish to recognize and acknowledge the very significant cultural role and reverence that the summit of Maunakea has always had within the Native Hawaiian community. We are most fortunate to have the opportunity to conduct observations from this mountain.

Guoshoujing Telescope (the Large Sky Area Multi-Object Fiber Spectroscopic Telescope LAMOST) is a National Major Scientific Project built by the Chinese Academy of Sciences. Funding for the project has been provided by the National Development and Reform Commission. LAMOST is operated and managed by the National Astronomical Observatories, Chinese Academy of Sciences.

\facilities{Gaia, Magellan: Clay (MIKE), Magellan: Baade (IMACS), APF, Max Planck: 2.2m, Keck: I (HIRES)}

\software{astropy \citep{astropy}, dustmaps3d \citep{dustmaps3d}, emcee \citep{fm13}, matplotlib
\citep{Hunter:2007}, numpy \citep{van2011numpy}, TheJoker \citep{joker}}

\bibliographystyle{apj}


\end{document}